\documentclass[12pt,draftclsnofoot,onecolumn]{IEEEtran}
\usepackage[table]{xcolor}
\usepackage{amsfonts}
\usepackage{filecontents}
\usepackage{amsmath}
\usepackage{tikz}
\usepackage{breqn}
\usepackage{graphicx}
\usepackage{pgfplots}

\usepackage{multirow}
\usepackage{booktabs}
\usepackage{subcaption}
\usepackage[]{algorithm2e}

\usetikzlibrary{shapes,decorations.pathmorphing,backgrounds,positioning,fit}
\usepackage{forest,array}
\usetikzlibrary{shadows}
\let\labelindent\relax
\usepackage{enumitem}

\def\paren#1{\left({#1}\right)}
\def\mean#1{E\left[  {#1}\right] }
\def\set#1{\left\lbrace {#1}\right\rbrace }
\DeclareMathAlphabet\mathbfcal{OMS}{cmsy}{b}{n}

\renewcommand\thesection{\Roman{section}}
\renewcommand\thesubsectiondis{\Alph{subsection}.}
\renewcommand\thesubsubsectiondis{\thesubsectiondis\arabic{subsubsection}.\quad}

\begin{document}

\title{DCSM Protocol for Content Transfer in Deep Space Network}
\author{\IEEEauthorblockN{Rojina Adhikary, John N. Daigle, and Lei Cao\\}
}

\maketitle
\begin{abstract}
		To maximize file transfer from space vehicles to the Earth, we propose a new space-to-earth content transfer protocol that combines turbo codes, RaptorQ codes, real-time channel prediction, and dynamic code-rate selection. The protocol features a practical signal-to-noise ratio prediction model that facilitates periodic adjustment of the turbo encoder to achieve adaptive rate transmission. Our simulation results indicate that an increase of about 20\% in file transfer rate is achievable using the proposed protocol.
	
\end{abstract}


\section{Introduction}\label{Introduction}
Every deep space mission has a communications system to receive commands and other information sent from Earth to the spacecraft and to return scientific data from the spacecraft to Earth. Without a consistently effective and efficient communications system, a successful mission would be impossible \cite{JPL}. In the case of the Mars missions, telemetry images and data files prepared by the Mars rovers are telemetered to the deep space network (DSN) stations on a daily basis. These files are reliably transmitted to an orbiter during a scheduled communication pass between the rover and the orbiter using the Proximity-1 protocol \cite{Secretariat2013}. These files are then safely stored in the orbiter memory and finally relayed to the DSN stations during scheduled communication passes between the orbiter and a DSN station.

The deep space communication channel over which content transfer takes place is characterized by large round-trip times (RTTs), intermittent connectivity, and highly variable propagation channels. For example, the RTT between Earth and Mars ranges from approximately 6.5 to 44 minutes as their interplanetary distance varies from approximately 60 million km to 400 million km. This large RTT renders the use of traditional content delivery protocols such as TCP/IP inefficient because in TCP, each unsuccessfully received packet invokes retransmission of the same packet with a delay on the order of an RTT.  In addition, signal degradation is inevitable due to the loss of signal energy with the distance and the thermal noise in the receiving system. Particularly, terrestrial and space weather variation causes continuous changes in atmospheric attenuation and the noise temperature of the DSN station receiver antenna system, which in turn varies the received signal-to-noise ratio (SNR). In future Mars exploration programs, content transfer will take place in the K-band and Ka-band where weather effects are larger than at X-band, which is used in current missions \cite{Laboratory2000}.  At present, the Mars Reconnaissance Orbiter (MRO) supports both the X-band and Ka-band communications and the Mars Odyssey (ODY) supports X-band only.

In the current deep space communication system, real time channel condition prediction is not utilized for content transfer from space vehicles to DSN stations. Instead, the standard method is to perform \emph{background sequencing} in which the spacecraft is commanded once every four weeks to reconfigure its transmission parameters. The use of \emph{mini-sequencing} to vary telecommunication parameters such as modulation index and data rate profile parameters of distant spacecraft on a weekly basis for Ka-band demonstration was introduced with the MRO \cite{Shambayati2005,Shambayati2007}. 

Using a pre-determined, fixed rate transmission scheme results in disruption of data continuity when bad weather occurs. In \cite{Sun2006}, an adaptive rate transmission scheme to combat the weather effect in the Ka-band link was proposed to maintain data continuity and high throughput. The work, however, has not considered coding in data delivery and any {\em a-priori} information in channel prediction. 

Existing research in delay tolerant networking (DTN) architecture with bundle protocol (BP) and Licklider transmission protocol (LTP) have shown that the utilization efficiency of link bandwidth can be affected by changing packet sizes at the bundle layer and the convergence layer. Along the same direction, in \cite{Lu2015}, a goodput enhancement algorithm (GEA) is proposed to find the optimal packet sizes in order to maximize goodput in one-hop and multi-hop DTNs. The work concludes that, at each hop, the optimal segment size is determined by the channel condition.

In this paper, we propose a new content delivery protocol specifically suitable for deep space communications. This protocol, termed as  dynamic code selection method (DCSM), consists of four key components: turbo codes \cite{Berrou1996}, RaptorQ codes \cite{RFC6330,RaptorQmonograph}, a channel prediction model, and a dynamic turbo code rate selection mechanism. This work is an extension of our preliminary work presented in \cite{Adhikary2017}.  
Turbo codes have been specified by the Consultative Committee for Space Data Systems (CCSDS). They are operated in the physical layer to correct corrupted bits at the receiver end. Utilizing our channel prediction model, turbo code rates of the spacecraft transmitter will be dynamically adjusted to provide variable packet size suitable for channel conditions as in \cite{Lu2015}. It should be noted that turbo codes could be substituted with any other advanced forward error correction (FEC) codes in the protocol. 

RaptorQ codes are the most advanced type of fountain codes that will be operated in the application layer on the application data units (ADUs) to reconstruct original files from received encoded symbols. Using RaptorQ codes, the source data symbols (or packets) can be encoded into a very large number of RaptorQ coded symbols. The source data symbols can be completely recovered if any subset of these coded symbols are received with the cardinality slightly larger than the number of original data symbols. In other words, the successful data recovery only depends on the number of encoded symbols that are correctly received, but does not depend on any specific symbols. As a result, the use of RaptorQ codes eliminates the need for the retransmission of any specific packets. 

Furthermore, to fully utilize the power of error correction codes and to minimize the effect of weather degradation in higher frequency band links, namely Ka and above, we propose a simple and practical channel bit-SNR prediction model based on a first order auto-regressive process AR(1). The prediction model also utilizes the {\em a-priori} information, which includes Mars-Earth geometry, pass duration and antenna elevation angles.
Given that the most important factors affecting transmission quality are measurable at a DSN station and significant computing resources are also available at the DSN station, weather conditions at the DSN station one RTT into the future can be predicted using both the {\em a-priori} information and real-time channel conditions. Based on the prediction result, the turbo encoder, including the code rate $r$ and block length $K$ bits, that maximizes the overall throughput of the channel one RTT in the future can be decided. The spacecraft is then commanded periodically with real-time, non-interactive commands to reconfigure its turbo encoder. 

We demonstrate the effectiveness of DCSM in comparison with two other methods: {\em genie} and  {\em static}. The {\em genie} method represents an ideal scenario in which it is assumed that the spacecraft knows the exact channel conditions arriving at a DSN station in advance so that it can adjust the turbo encoder to achieve the maximum link capacity and hence provides an upper-bound in performance of a turbo based adaptive rate transmission scheme. In the {\em static} method, a turbo encoder with fixed block length and code rate is used throughout the duration of a communication pass. This method mimics the current approach used for X-band communication with the Mars Odyssey. Although two different rates per pass are currently used for the MRO mission with 8920-bit information block length turbo codes and an algorithm called DR-90 employed \cite{Shambayati2005},  the data-rate profile for the allocated passes is determined at the beginning of each 28-day sequencing period.

For these three protocols, we divide them into a few sub-classes with different specifications and study their effects. For example, we have tested the effect of transmitting different numbers of symbols based on different estimation methods in the RaptorQ codes. 
To balance between maximizing the channel throughput and minimizing the amount of time required to deliver each file, we consider different arrangements for file transmissions. These alternative arrangements include transmitting files strictly serially or interleaved and RaptorQ encoding of individual files  or groups of files.  We will show that the method of serial transmission of separately RaptorQ encoded files along with proactive transmission of additional encoded symbols stands out and should be used as the practical implementation of the DCSM protocol. Using practical channel conditions in space communications, we demonstrate through simulations that the proposed DCSM protocol can achieve 99.9\% of the {\em genie}'s performance in throughput and yields an improvement of about 20\% over the {\em static} method. This suggests that the DCSM protocol can be a good candidate for future Ka and above frequency band space communications that involves significant RTT, noise levels and weather degradation effects. 

The rest of the paper is organized as follows. Section~\ref{sec:system_model} describes the overall system, proposed protocol stack, turbo encoder selection approach, file arrangement  and a data loss mitigation analysis.  In Section~\ref{Section:Elaborations_to_the_Methods}, the DCSM, {\em genie} and {\em static} methods are described together with their sub-classes. In Section~\ref{sec:noise_temperature_bit_SNR}, the details of calculating channel bit-SNR starting from wet path delay information are presented. Section~\ref{sec:prediction_model} presents the channel bit-SNR prediction model. Section~\ref{sec:simulation} presents communication data rates corresponding to the selection of turbo encoder and compares, via simulation, the proposed DCSM method with the upper bound given by the {\em genie} method and the {\em static} method for the MRO Ka-band communication scenario. Section~\ref{sec:conclusions} presents conclusions.

\section{System Model}
\label{sec:system_model}

\begin{figure}[h!]
	\centering
	\begin{tikzpicture}[scale=0.95]
	\begin{scope}
	\begin{scriptsize}
	\draw [thick] (2.5,0) rectangle (6.5,4) node[midway, text width=2cm, align=center]{Receive real-time turbo code selection commands and extract the specified turbo code};
	\draw[-latex] (6.5,2) -- (7.5,2);
	\draw [thick] (7.5,0) rectangle (11.5,4) node[midway, text width=2cm, align=center]{Transmit RaptorQ $+$ turbo encoded (using the specified code rate and block length) telemetry data};
	\draw[latex-latex] (2.5,-0.6) -- (11.5,-0.6) node[midway,fill=white] {Mars orbit relay spacecraft at time $(t_0 + \mathrm{RTT}/2 )$};			
	\draw[-latex] (11.5,2) -- (15,2) -- (15,-5) node[midway,fill=white,rotate= -90] {Reverse channel} -- (14,-5);
	\draw [thick] (10,-7) rectangle (14,-3) node[midway, text width=2cm, align=center]{Receive telemetry data packets and perform RaptorQ $+$ turbo decoding of the received data packets};
	\draw[latex-] (9,-5) -- (10,-5);
	\draw [thick,fill=gray!45] (5,-7) rectangle (9,-3) node[midway, text width=2cm, align=center]{Predict channel bit-SNR at time $(t_0 + \mathrm{RTT})$ and generate real-time turbo selection command};
	\draw[latex-] (4,-5) -- (5,-5);
	\draw [thick] (0,-7) rectangle (4,-3) node[midway, text width=2cm, align=center]{Transmit the real time turbo code (code rate and block length) selection command to the orbiter};
	\draw[latex-latex] (0,-7.6) -- (14,-7.6) node[midway,fill=white] {DSN station at time $t_0$};
	\draw[-latex] (0,-5) -- (-1,-5) -- (-1,2) node[midway,fill=white,rotate= 90] {Forward channel} -- (0,2) -- (2.5,2) ;
	\end{scriptsize}
	\end{scope}
	\end{tikzpicture}
	\caption{Summary of encoding and decoding processes occurring between a DSN station and the orbiter.}
	\label{fig:process}
	\vspace*{-\baselineskip}
\end{figure}
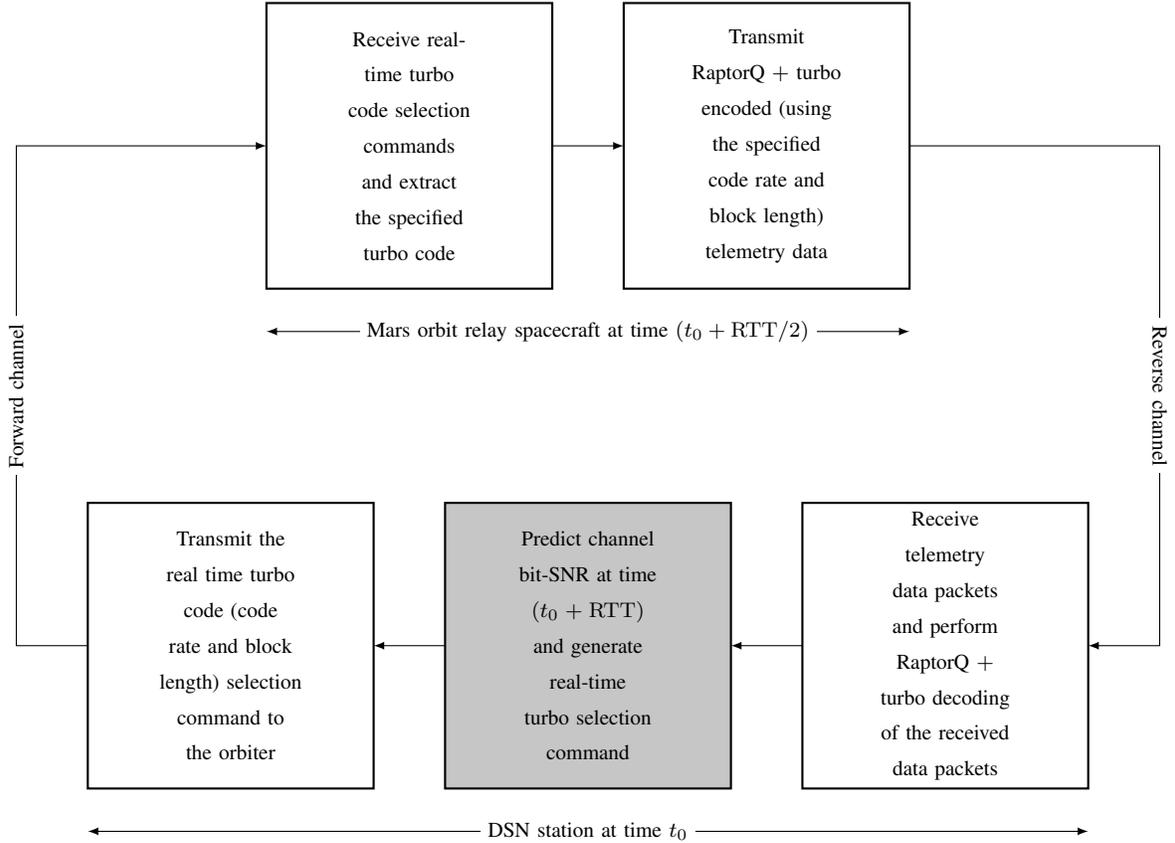

The communication diagram of the DCSM protocol is shown in Fig.~\ref{fig:process}. Two separate processes are simultaneously occurring at a communicating DSN station: a) channel condition prediction, appropriate turbo code selection and command generation, and real-time commanding of the orbiter, and b) telemetry data reception and decoding. Similarly, two separate processes are simultaneously occurring at the orbiter: a) real time command reception, decoding, and appropriate turbo encoder selection and b) RaptorQ plus turbo encoding and telemetry data transmission. 

The channel condition prediction algorithm is continuously executed at the DSN station to predict channel bit-SNR one RTT into the future. The turbo code that maximizes throughput for the predicted channel condition is decided, and real time turbo encoder selection commands are generated and transmitted to the orbiter. The orbiter transmits turbo encoded symbols from current file one-at-a-time in succession, and then starts transmitting encoded symbols for the next file. The DSN collects correctly received RaptorQ encoded symbols from the file and periodically reports the number of additional encoded symbols to be transmitted from the orbiter to ensure successful decoding of the file. The orbiter responds by promptly transmitting the specified number of additional encoded symbols.

\subsection{Protocol stack: RaptorQ codes over Turbo codes}
\label{sec:protocol_stack}
In ODY, the telemetry signal is encoded with two codes, a Reed-Solomon $\paren{255,223}$ as the outer code and a $\paren{7,1/2}$ convolutional code as the inner code \cite{AndreMakovskyOctober}. In MRO, three different coding types are available \cite{Shambayati2007}, \cite{JimTaylorSeptember2006}: a) $\paren{255,223}$ Reed-Solomon block code with interleaving depth of either 5 or 1, b) $\paren{255,223}$ Reed-Solomon and a $\paren{7,1/2}$ convolutional code with interleaving depth of 5, and c) turbo codes with block length 8920 bits and rates $1/2$, $1/3$, and $1/6$. 

In this paper, we propose the use of turbo codes in the physical layer and the RaptorQ class of fountain codes in the application layer for the data telemetry from orbiters to DSN stations. Turbo codes are one of the most advanced FEC codes that can achieve near-Shannon-limit error correction performance with reasonable coding and decoding complexity \cite{Berrou1996}, \cite{Berrou1993}. Good turbo codes can come within approximately 0.8 dB of the theoretical limit at a bit error rate (BER) of $10^{-6}$. With turbo codes, synchronization is accomplished by preceding each transfer frame with a rate dependent attached synchronization marker. For the CCSDS recommended codes, the turbo decoder error floor occurs at a BER of less than $10^{-7}$. For operation near this region, CCSDS recommends (optional) that a 16 bit cyclic redundancy check (CRC) be inserted at the end of the codeblock as an independent check on the
decoding process \cite{Laboratory2000,G-22012,B-22011}. 

RaptorQ codes are the most advanced and efficient fountain codes designed to date \cite{RaptorQmonograph}, \cite{Shokrollahi2006}. RaptorQ codes can generate very large numbers of encoded symbols but the receiver needs only a specific number of unique encoded symbols in order to recover the original content. As long as the receiver is able to collect a number of encoded symbols slightly more than the number of original source symbols, it can successfully recover the original file. The success of recovery only depends on the number of encoded symbols received, but not on any specific symbols. As a result, the RaptorQ-based protocol removes the need of negative-acknowledgment (NAK) for any specific symbol and hence largely improves the performance of the delay tolerant content transfer.

	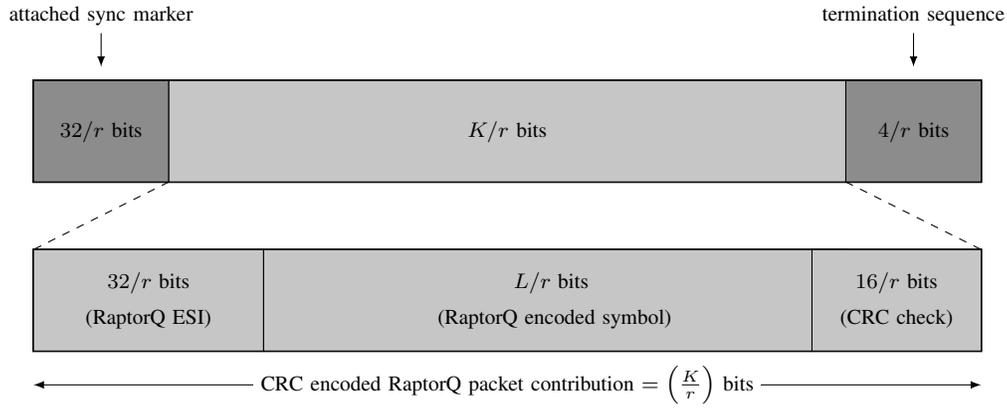
\begin{figure}[h!]
		\centering
		\begin{tikzpicture}[scale=0.9]
		\begin{scope}
		\begin{scriptsize}
		\draw[latex-] (1,1.7) -- (1,2.2) node[above] {attached sync marker};
		\draw [thick] (0,0) rectangle (14,1.5) ; 
		\draw[fill=gray!90] (0,0) rectangle (2,1.5) node[midway, color=black] {$32/r$ bits};
		\draw[fill=gray!45] (2,0) rectangle (12,1.5) node[midway,color=black] {$K/r$ bits};
		\draw[fill=gray!90] (12,0) rectangle (14,1.5) node[midway, color=black] {$4/r$ bits};
		\draw[latex-] (13,1.7) -- (13,2.2) node[above] {termination sequence};
		\draw[latex-latex] (0,-3) -- (14,-3) node[midway,fill=white] {CRC encoded RaptorQ packet contribution $= \paren{\frac{K}{r}}$ bits};
		\draw[dashed] (2,0) -- (0,-1);
		\draw[dashed] (12,0) -- (14,-1);
		\draw [thick] (0,-2.5) rectangle (14,-1) ; 
		\draw[fill=gray!45] (0,-2.5) rectangle (3.4,-1) node[midway, color=black, text width=2cm, align=center] {$32/r$ bits (RaptorQ ESI)};
		\draw[fill=gray!45] (3.4,-2.5) rectangle (12,-1) node[midway,color=black, text width=4cm, align=center] { $\hspace{2 em} L/r$ bits $\hspace{2 em}$ (RaptorQ encoded symbol)};
		\draw[fill=gray!45] (11.5,-2.5) rectangle (14,-1) node[midway,color=black, text width=2cm, align=center] {$16/r$ bits (CRC check)};
		\end{scriptsize}
		\end{scope}
		\end{tikzpicture}
		\caption{Turbo encoded frame with a RaptorQ symbol.}
		\label{fig:turbo_codeword}
		\vspace*{-\baselineskip}
	\end{figure}

The RaptorQ encoder treats each file to be telemetered to DSN as a source block and divides it into $K_{\text{S}}$ source symbols of size $L$ bits. These source symbols are then RaptorQ encoded to generate $N_{\text{E}}$ encoded symbols of same size, where $K_{\text{S}} < N_{\text{E}}$. Each encoded symbol has an associated encoded symbol identifier (ESI) of size 32 bits to uniquely identify the symbol. A RaptorQ encoded symbol along with its ESI forms a RaptorQ packet with length of  $L + 32$ bits. 
Each RaptorQ packet is an application protocol data unit (APDU) that is further handed over to the CRC encoder to create a CRC encoded transfer frame of length $K = L + 32 + 16$ bits and the transfer frame is turbo encoded before transmission. The length of a turbo encoded frame is hence $(K + 36)/r$ bits. The frame structure is shown in Fig.~\ref{fig:turbo_codeword}. As one packet has one RaptorQ coded symbol, ``packet" and ``symbol" will be used interchangeably in the sequel. 

At the receiver, a clean RaptorQ packet is extracted after turbo decoding and handed over to the RaptorQ decoder. Once the number of RaptorQ encoded symbols received by the RaptorQ decoder is sufficient, the original file is reconstructed. The RaptorQ decoder requires $\paren{K_{\text{S}} + \Theta }$ symbols to be able to successfully decode a source block with $K_{\text{S}}$ source symbols, where $\Theta \in \set{0, 1, 2, \cdots}$ is symbol overhead. As shown in \cite{RaptorQmonograph}, with $\Theta = 0$, the source block can be successfully decoded with probability of 99\%; with $\Theta = 1$ and $\Theta = 2$, the source block is successfully decoded with probability of 99.99\% and 99.9999\%, respectively. The higher the symbol overhead, the higher is the probability of successful decoding.

\subsection{Turbo encoder selection}
\label{sec:turbo_encoder_selection}
The computed channel bit-SNR can be translated into frame error rates (FERs) for all alternative turbo block sizes $K$ and code rates $r$. In the DCSM scheme, the turbo code is selected to yield the highest data reception rate for the channel bit-SNR value. Unless stated otherwise, the CRC overhead is not included in the following as its implementation is optional. 

Let $P_E(\text{SNR}, K_i, r_j)$ be the FER corresponding to the use of turbo encoder with information block length $K_i \in \set{1784, 3568, 7136, 8920}$ bits and code rate $r_j \in \set{1/2, 1/3, 1/4, 1/6}$ at the SNR value. The rate of packet transmission from the orbiter with gross bit transmission rate of $R_b$ bps is $ R_b \, r_j/ (K_i + 36)$ packets/s. The rate at which correct packets are received at the DSN is $ R_b \, r_j \, \paren{1 - P_E\paren{\text{SNR}, K_i, r_j}  }/ (K_i + 36)$ packets/s. Overall throughput is thus $ R_b \, r_j \, \paren{1 - P_E\paren{\text{SNR}, K_i, r_j}}L_i/(K_i+36)$ bits/s. Mathematically, the selection of turbo encoder can be obtained by maximizing the overall throughput, i.e.,
\begin{align}
\label{eq:maximize}
\quad\quad\quad\quad \max_{i,j} \,\,  \, \paren{1 - P_E\paren{\text{SNR}, K_i, r_j}  } \, r_j\frac{K_i-32}{K_i+36}.
\end{align}

\begin{table}[h!]
	\begin{center}
		\caption{Dynamic turbo code assignment for different channel bit-SNR ranges and their names.}
		\begin{tabular}{l|@{\qquad}c|@{\qquad}c}\toprule
	Name & Channel bit-SNR (dB) range & turbo code $(K, r_i)$\\ \toprule
	Range A & $< -0.5$ & - \\
	Range B & $-0.5 \sim -0.1$ & $(8920,1/6)$ \\
	Range C & $ -0.1 \sim 0.15$ &  $(8920,1/4)$ \\
	Range D & $0.15 \sim 0.85$  &  $(8920,1/3)$ \\
	Range E & $0.85 \sim 2.2$ & $(8920,1/2)$ \\
	\bottomrule
\end{tabular}
		\label{table:dynamic_assignment}
	\end{center}
		\vspace*{-\baselineskip}
\end{table}

Using \eqref{eq:maximize} against the performance of turbo codes presented in \cite{Laboratory2000} and \cite{G-22012}, we obtain the dynamic code assignments shown in Table~\ref{table:dynamic_assignment} for achieving the highest link throughput. Out of the sixteen turbo code options recommended by CCSDS, under different channel conditions, link throughput is maximized when $K=8920$ bits. This result is in line with what is expected with turbo codes because they always perform better with a larger packet size \cite{Berrou1996}. Thus, with 4 possible rates the real-time commands generated at the DSN station needs only two bits to reflect the selected code rate $r$. 

The result also signifies that as long as the channel bit-SNR prediction model is capable of predicting the correct bit-SNR range, optimal turbo encoder to be used remains the same. Thus, although we continuously predict bit-SNR value one RTT into the future, whenever our predicted bit-SNR falls in the same range where the actual bit-SNR resides, the prediction is considered as correct prediction.

As long as the turbo encoder to be used by the orbiter remains the same, there is no need for the DSN to transmit the rate adjustment command continuously. 
To minimize the use of uplink, the code rate selection command is generated only when the prediction model detects the future bit-SNR moving from one range to another. 
In the case where predicted channel bit-SNR is below the communication threshold, i.e., $ < -0.5$ dB, indicating weather dropout conditions, the orbiter is commanded to stop transmission until it receives a clear to send (CTS) command when a better bit-SNR is predicted.

\subsection{File arrangement}
\label{sec:file_arrangement}

Considering delivery time of individual source files and possible throughput, there are a few ways of using RaptorQ codes and file arrangement for data transmission.  
In this paper, the method used is the serial transmission of separately RaptorQ encoded files along with proactive transmission of additional encoded symbols.  Each file is RaptorQ coded independently and transmitted sequentially. For example, the $\paren{K_{\text{S}} + \Theta + y_m }$ symbols from the $m$-th file are transmitted, where $y_m$ is the number of additional encoded symbols used to protect against packet loss due to unsuccessful turbo decoding. This is followed by immediately transmission of the ($m$+1)-st file. At the receiver end, as soon as the number of corrupted RaptorQ symbols received for the $m$-th file exceeds $y_m$, feedback message(s) requesting transmissions of additional encoded symbols of the $m$-th file is sent over to the spacecraft each time the receiver knows it has a symbol loss. As soon as a feedback is received at the spacecraft, transmission of symbols from the current file is temporarily paused until the requested number of symbols from the $m$-th file are transmitted. Using this approach, we target to deliver each file as soon as possible with reasonable processing and memory usage. It is essential to note that uplink is used for real-time turbo encoder selection commands as well as for feedbacks requesting additional number of symbols from previously transmitted files. The feedback requests can be piggybacked on real-time turbo encoder selection commands. 

\begin{figure}[h!]
	\centering
	\includegraphics[width=5.4in]{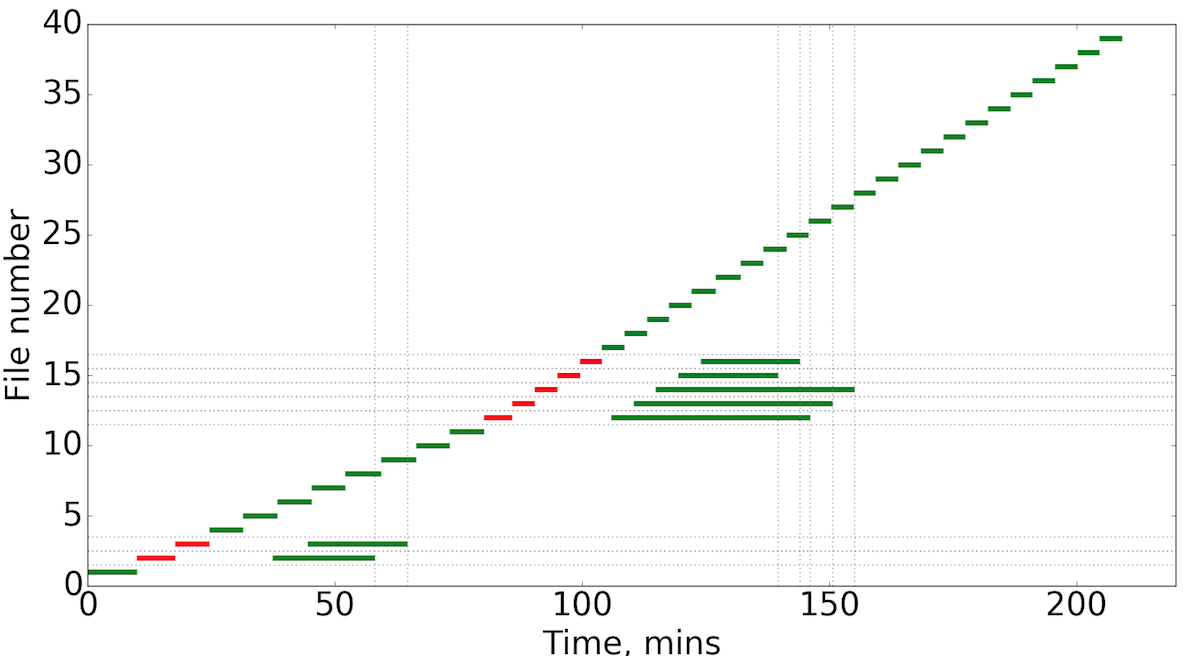}
	\caption{An example output showing serial transmissions of files during a communication pass.}
	\label{fig:communication_pass_March_13_2008}
	\vspace*{-\baselineskip}
\end{figure}

One practical process with the proposed system for 20 min RTT delay is presented in Fig.~\ref{fig:communication_pass_March_13_2008}. The horizontal red lines indicate that the number of clean packets received were not enough to decode the file and additional transmissions were needed. The horizontal green lines indicate successful decoding of a file and its length equals the total time required for the file transmission. It can be observed from the figure that the first file was successfully recovered in the first round of  transmission. For the 2nd file, symbols received during the first round were insufficient and additional transmissions were requested. Additional transmission began where its corresponding green line began, and took some RTTs to receive enough symbols to successfully decode the file. It is also worth mentioning that each retransmission request is not for requesting any specific symbols. It only tells the transmitter how many more symbols should be transmitted. 

Other file arrangement methods might be possible too. For example, one is file interleaving where each file is RaptorQ coded but encoded symbols from a number of $Q$ files are interleaved together. We use $Q$ transmission queues to hold encoded symbols from $Q$ interleaved files at a time. A proportion of symbols from each interleaved file are transmitted at a time until the required number of symbols from all the $Q$ files are transmitted. Once the first $Q$ files are done, another set of $Q$ files are interleaved for transmission and the process goes on. Interleaving results in re-distribution of errors among multiple files in case of burst errors, and one file does not have to bear all the losses. However, it requires more standby queues and has larger memory consumption and complexity. It might also lead to large number of incomplete files in the system at a given time (a portion of multiple received files are missing), and result in increased delivery time for each file. Another file arrangement method could be that instead of RaptorQ encoding each file individually, all the files to be transmitted are RaptorQ encoded at once, and then transmitted. Although this method may result in higher throughput and has lowest complexity in comparison with the above methods, real time delivery of each file is impossible as all files have to be re-constructed simultaneously at the end of transmission. This is particularly undesired if there are multiple priority classes.
	
\subsection{Data loss and mitigation approach}
\label{sec:y_calculation}
Space communications are characterized by noisy channels with a BER of $10^{-5}$ being very common and even higher BERs on the order of $10^{-1}$ in the deep space environment \cite{Akyildiz:2003}. Although powerful turbo codes are used, unsuccessful decoding is inevitable, and in turn causes RaptorQ symbol loss at the receiver. Thus, it is beneficial to transmit $\paren{K_{\text{S}} + \Theta + y }$ encoded symbols from a file initially instead of $\paren{K_{\text{S}} + \Theta }$ symbols, where $y$ is the additional number of symbols to be transmitted to counter the effect of possible transmission failures so that the number of re-transmissions required is kept to the minimum. This strong capability of being able to transmit additional symbols proactively comes from the use of RaptorQ codes. 


During a communication pass, for each symbol (turbo frame) transmission, indexed by $n$, there is an associated turbo code rate $r^{(n)}$ and FER $p^{(n)}$ that maximizes the channel throughput at the bit-SNR. Even though the bit-SNRs are correlated, the probability of a frame error $p^{(n)}$ at a bit-SNR value depends specifically upon the specific realization over the period that the frame is being transmitted. The realizations are drawn independently so the frame errors are independent. Thus, each symbol transmission is an independent trial. Let $\tilde{y}_n$ be a random variable whose value is equal to 1 if trial $n$ is a failure, and 0 otherwise, represented as
\begin{equation}
	\nonumber \tilde{y}_n = \left\{
	\begin{array}{l l}
		1 & \quad \text{failure, }\\
		0 & \quad \text{otherwise.}
	\end{array} \right.
\end{equation}
It follows that the random number of failures $\tilde{y}$ out of the $\paren{K_{\text{S}} + \Theta + y}$ transmitted symbols of a file can be expressed as
$\tilde{y} = \sum_{n=1}^{\paren{K_{\text{S}} + \Theta + y}} \tilde{y}_n.$
The $\tilde{y}$ is approximately Gaussian distributed as 
\begin{equation}
	\tilde{y} \sim \mathcal{N} \paren{\sum_{n=1}^{\paren{K_{\text{S}} + \Theta + y}} p^{(n)}, \sum_{n=1}^{(K_{\text{S}} + \Theta + y)} p^{(n)} (1-p^{(n)}) }.
	\label{eqn:y}
\end{equation}
We now compute the $y$ additional symbols to be transmitted so that the future packet losses of that file do not affect the recovery of the source symbols. To guarantee this, the value of $y$ should be chosen such that $P\set{\tilde{y} \leq y}$ is greater than or equal to some large  target probability $p_{\mathrm{target}}$. That is,
\begin{align}
	P\set{\tilde{y} \leq y} \geq p_{\mathrm{target}},  \notag\\
	P\set{\frac{\tilde{y} - \mean{\tilde{y}}}{\sqrt{\operatorname{Var}\paren{\tilde{y}}}} > \frac{y - \mean{\tilde{y}}}{\sqrt{\operatorname{Var}\paren{\tilde{y}}}}} \leq 1 - p_{\mathrm{target}},  \notag \\ 
	y \geq \sqrt{\operatorname{Var}\paren{\tilde{y}}} \operatorname{Q}^{-1}\paren{1 - p_{\mathrm{target}}} + \mean{\tilde{y}},
	\label{eqn:y_value}
\end{align}
where $Q(z)=\int_z^\infty\frac{1}{\sqrt{2\pi}}e^{-\frac{x^2}{2}}dx $ is the right tail function of the standard normal distribution. 
Basically, we can proceed in two different ways with computing and using $y$ during a communication pass; a) fixed approach, and b) adaptive approach. Let the value of $y$ for file $m$ be by $y_m$, $m =1, 2, 3, \cdots, F_{\mathrm{tx}}$, where $F_{\mathrm{tx}}$ is the total number of files transmitted in the pass. In the \emph{fixed approach}, $y_1 = y_2 = \cdots = y_{ F_{\mathrm{tx}} } = y$. At the beginning of each pass, the same $y$ is computed at the DSN station utilizing the previous communication pass information, and relayed to the spacecraft. In the \emph{adaptive approach}, $y$ is computed for each file separately at the spacecraft, i.e., $y_1, y_2, \cdots, y_{ F_{\mathrm{tx}} } $, on the basis of proportion of time of different turbo encoders are used for transmission of symbols from the file and its corresponding FERs.

\section{Protocols}
\label{Section:Elaborations_to_the_Methods}

In this section, we describe the protocols and  sub-classes of the DCSM, the \emph{genie} and the \emph{static} approaches, as shown in Fig. 4. Mathematical formulation of channel throughput and computation of $y_m$ are also provided. 

\newcolumntype{C}[1]{>{\centering}p{#1}}
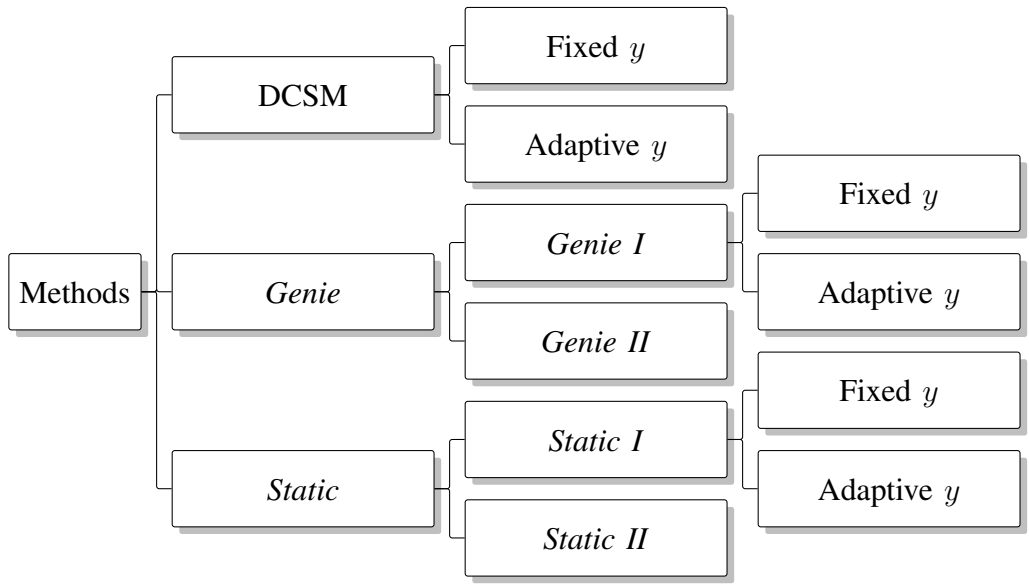
\begin{figure}[h!]
	\centering   
	\begin{forest}
		for tree={
			if level=0{align=center}{
				align={@{}C{32mm}@{}},
			},
			grow=east,
			draw,
			edge path={
				\noexpand\path [draw, \forestoption{edge}] (!u.parent anchor) -- +(2mm,0) |- (.child anchor)\forestoption{edge label};
			},
			parent anchor=east,
			child anchor=west,
			l sep=4mm,
			tier/.wrap pgfmath arg={tier #1}{level()},
			edge={ rounded corners=1pt},
			fill=white,
			rounded corners=1pt,
			drop shadow,
		}
		[Methods
		[\emph{Static}
		[\emph{Static II}]
		[\emph{Static I}
		[Adaptive $y$]
		[Fixed $y$]
		]
		]	
		[\emph{Genie}
		[\emph{Genie II}]
		[\emph{Genie I}	
		[Adaptive $y$]
		[Fixed $y$]
		]
		]
		[DCSM
		[Adaptive $y$]
		[Fixed $y$]
		]
		]
		]
	\end{forest}
	\label{fig:myforest}
	\caption{Tree representing all the methods and sub-categories.}
	\vspace*{-\baselineskip}
		\vspace*{-\baselineskip}
\end{figure}

\subsection{DCSM}


Features of the DCSM approach include a) RaptorQ codes at the application layer, b) dynamic selection of turbo codes at the physical layer, c) a channel condition prediction model to predict channel bit-SNR one RTT into the future, d) $y_m$ additional symbols transmitted for the $m$-th file and e) real-time feedback on the additional number of symbols to be transmitted from a file for its successful recovery. A stepwise elaboration on execution of the DCSM is given below:
\begin{enumerate}[label=\roman*)]
	\item At the spacecraft, for the $m$-th file to be telemetered to the Earth station, $\paren{K_{\text{S}} + \Theta + y_m}$ RaptorQ encoded symbols are generated, and handed over to the physical layer for turbo encoding.
	\item An appropriate turbo encoder is used to encode each RaptorQ encoded symbol as summarized below:
	\begin{enumerate}
		\item At the Earth station, at time $t_0$ a prediction of future channel bit-SNR $X_{t_0 + \mathrm{RTT}}$ that will occur at time $t_0 + \mathrm{RTT}$ is made.
		\item A real-time command identifying the turbo encoder that maximizes channel throughput at time $t_0 + \mathrm{RTT}$ is sent over to the spacecraft.
		\item The spacecraft receives the real-time command at time $t_0 + \mathrm{RTT}/2$ and uses the specified turbo encoder. 
	\end{enumerate}
	\item The Earth station keeps a record of the number of packets successfully received as well as packets lost from a file. This information is easily extracted by comparing the number of successfully received packets and the ESI of the RaptorQ encoded symbol contained in symbols. 
	\item Once the number of lost packets from file $m$ hits $y_m$, for every new symbol loss recorded, feedback(s) requesting transmission of that number of additional symbols is sent over to the spacecraft. A feedback can be for one additional packet as well as for multiple additional symbols.
	 \item The spacecraft continues with the transmission of $\paren{K_{\text{S}} + \Theta + y_m}$ symbols from each file successively. The instant it receives a feedback, current transmission is paused temporarily until the specified number of symbols from the specified file is generated and transmitted. 
	\item As soon as the Earth station receives $\paren{K_{\text{S}} + \Theta}$ packets successfully, the file is successfully decoded and no further feedback for this file will be created.
\end{enumerate}

During a communication pass, let $\alpha_i$ be the proportion of time during which the actual channel bit-SNR is such that the packets are transmitted at rate $r_i$, $r_i \in \set{0, 1/2, 1/3, 1/4, 1/6}$. The $\alpha_i$ is known to \emph{genie} only. Rate $r_i = 0$ represents the weather dropout conditions at the Earth station. 

Let $\beta_i$ be the proportion of time during which the channel condition prediction model predicts a bit-SNR such that the packets are transmitted at rate $r_i$. If the prediction method is highly accurate, then $\beta_i \approx \alpha_i$. Thus, with the DCSM approach, over a communication pass of duration $T$ seconds, we spend $T'_i = \beta_i \, T$ seconds transmitting at rate $r_i$. However, for maximizing throughput it should have been $T_i = \alpha_i \, T$ seconds.

From Fig.~\ref{fig:turbo_codeword}, the length $L_p^{(r_i)}$ of a rate $r_i$ turbo encoded packet is 
\begin{equation}
L_p^{(r_i)} = (36 + K)/r_i \quad \mathrm{bits.}
\label{eqn:pkt_length}
\end{equation}  
Data transmission rate $R_d^{(r_i)}$ for code rate $r_i$ turbo codes is related to the bits transmission rate $R_b$ of the channel as
\begin{equation}
R_d^{(r_i)} = \frac{L}{L_p^{(r_i)}} \, R_b = \frac{(K-32)}{(K+36)} \, R_b \, r_i \quad \mathrm{bps}.
\label{eqn:data_rate}
\end{equation} 
Over a communication pass duration, the number of symbols (packets) transmitted $N_{\mathrm{tx}}^{(d)}$ is given as
\begin{equation}
N_{\mathrm{tx}}^{(d)} = \sum_{r_i \neq 0} r_i \, \beta_i \, T  \, R_b \, \frac{1}{(K+36)},
\end{equation}
where the superscript ``$(d)$" denotes DCSM.

For a communication pass, define
$p_{ij}$, $i, j \in \set{A, B, C, D, E}$, as the FER when bit-SNR range $i$ is predicted as range $j$ and the optimal turbo code corresponding to range $j$ is used for the channel with bit-SNR actually in range $i$. Clearly, $p_{ii}$ is the FER when bit-SNR rate is correctly predicted and the corresponding optimal turbo code is used. Out of the $\paren{K_{\text{S}} + \Theta + y_m}$ transmitted symbols of the $m$-th file, assume that the bit-SNR range is accurately predicted for $k_{\text{s}}$ transmitted symbols and the remaining $(K_{\text{S}} + \Theta + y_m - k_{\text{s}})$ symbols are transmitted when bit-SNR prediction is incorrect. The random number of failures, $\tilde{y}_m$, out of $\paren{K_{\text{S}} + \Theta + y_m}$ transmitted symbols of $m$-th file is Gaussian distributed with mean $\mean{\tilde{y}_m} = \paren{\sum_{n=1}^{k_{\text{S}}} p_{ii}^{(n)} + \sum_{n = k_{\text{S}} + 1}^{\paren{K_{\text{S}} + \Theta + y_m}} p_{ij}^{(n)}}$ and variance $\operatorname{Var}\paren{\tilde{y}_m} = \paren{\sum_{n=1}^{k_{\text{S}}} p_{ii}^{(n)} (1-p_{ii}^{(n)}) + \sum_{n = k_{\text{S}} + 1}^{\paren{K_{\text{S}} + \Theta + y_m}} p_{ij}^{(n)} (1-p_{ij}^{(n)})}$,
where superscript ``$(n)$" denotes the FER corresponding to the $n$-th transmitted symbol.
Theoretically, using~\eqref{eqn:y_value} along with the $\mean{\tilde{y}_m}$ and the $\operatorname{Var}\paren{\tilde{y}_m}$, the value of $y_m$ can be computed.

Similarly, the random number of failures $\tilde{Y}^{(d)}$ out of $N_{\mathrm{tx}}^{(d)}$ symbol transmissions  during a communication pass is also Gaussian distributed and can be represented as sum of $\tilde{y}_{m}$  
\begin{equation}
\tilde{Y}^{(d)} = \sum_{m = 1}^{F_{\mathrm{tx}}^{(d)}} \tilde{y}_{m},
\label{eqn:sum_2}
\end{equation}
where $F_{\mathrm{tx}}^{(d)}$ is the total files transmitted during a communication pass. Average channel throughput $T_{H}^{(d)}$, in terms of the number of correctly received symbols at the receiver per second is $T_{H}^{(d)} = (N_{\mathrm{tx}}^{(d)} - \tilde{Y}^{(d)})/T$. 

To compute the $y_m$ using~\eqref{eqn:y_value}, we first need $\operatorname{Var}\paren{\tilde{y}_m}$ and $\mean{\tilde{y}_m}$, for which exact knowledge of $p_{ii}^{(n)}$ and $p_{ij}^{(n)}$ is needed. This is impossible due to unknown prediction accuracy. Therefore, it is not feasible to compute the actual value of $y_m$ in a real implementation. To find approximate $y_m$, we propose two alternatives: fixed DCSM and adaptive DCSM.\\

\subsubsection{Fixed DCSM}
$\quad$\\*[-0.75em]

In this approach, we compute $y = y_1 = y_2 = \cdots = y_{ F_{\mathrm{tx}}^{(d)} }$ utilizing actual and the predicted bit-SNR profile of the previous communication pass. Using statistics of the past communication pass, we obtain statistics of proportion of time $\alpha_i$ bit-SNR is in a range $i$, proportion of time $\beta_{ij}$ bit-SNR is predicted to be in range $j$ when it is actually in range $i$, and total proportion of time $\beta_j = \sum_{i} \beta_{ij} $ bit-SNR is predicted in range $j$. Thus, $\beta_{ij}, i = j$ and $\beta_{ij}, i \neq j$ respectively denote correct and incorrect predictions of bit-SNR ranges. This can be summarized in tabular form as presented in Table~\ref{table:statistics}.

\begin{table}[h!]
	\caption{ Actual and predicted bit-SNR occurrence proportion of a communication pass.}
	\centering
	\begin{tabular}{c|c|c|c|c|c}
		\toprule
		\multirow{3}{*}{\raisebox{-\heavyrulewidth}{Proportion of SNRs}} & \multicolumn{5}{ c }{Predicted SNR range proportion} \\
		\cmidrule{2-6} 
		 & Range A  & Range B & Range C & Range D & Range E \\
		\midrule
		$\alpha_A$ & \cellcolor{gray!45}$\beta_{AA}$ & $\beta_{AB}$ & $\beta_{AC}$ & $\beta_{AD}$ & $\beta_{AE}$\\
		$\alpha_B$ & $\beta_{BA}$ & \cellcolor{gray!45}$\beta_{BB}$ & $\beta_{BC}$ & $\beta_{BD}$ & $\beta_{BE}$\\
		$\alpha_C$ & $\beta_{CA}$ & $\beta_{CB}$ & \cellcolor{gray!45}$\beta_{CC}$ & $\beta_{CD}$ & $\beta_{CE}$\\
		$\alpha_D$ & $\beta_{DA}$ & $\beta_{DB}$ & $\beta_{DC}$ & \cellcolor{gray!45}$\beta_{DD}$ & $\beta_{DE}$\\
		$\alpha_E$ & $\beta_{EA}$ & $\beta_{EB}$ & $\beta_{EC}$ & $\beta_{ED}$ & \cellcolor{gray!45}$\beta_{EE}$\\
		\midrule
        Proportion of pred. SNRs & $\beta_A$ & $\beta_B$ & $\beta_C$ & $\beta_D$ & $\beta_E$\\
		\bottomrule
	\end{tabular}
	\label{table:statistics}
	\vspace*{-\baselineskip}
\end{table}

Matrix form representation of the range prediction information is given by the range prediction matrix  $\boldsymbol{\beta}$ as
\begin{align}
\boldsymbol{\beta} = \left[ \beta_{ij} \right], \,\, \mathrm{for} \quad i,j \in \set{A, B, C, D, E}.
\label{eqn:matrix}
\end{align}

Let the minimum and maximum bit-SNRs corresponding to range $i$ be $\text{SNR}_{i_{\mathrm{min}}}$ and $\text{SNR}_{i_{\mathrm{max}}}$, respectively. Corresponding to each bit-SNR $\text{SNR}_{i}$ inside range $i$, there is an associated achievable FER value $P_E(\text{SNR}_{i})$. By averaging over all FERs in each range $i$, we obtain average FER corresponding to each range given as
\[
p_{i_{\mathrm{avg}}} = \frac{1}{\mathrm{card}(i)}\sum_{\text{SNR}_i=\text{SNR}_{i_{\mathrm{min}}}}^{\text{SNR}_{i_{\mathrm{max}}}} P_E(\text{SNR}_{i}),
\] 
where $\mathrm{card}(i)$ is the number of SNR points in range $i$. For range $i$, the average FER is $p_{i_{\mathrm{avg}}}$. However, due to the range estimation error, in reality, the average FER $\hat{p}_{i}$ experienced by the rate $r_i$ turbo encoded packets is different from $p_{i_{\mathrm{avg}}}$. Thus, utilizing $\boldsymbol{\beta}$ of past communication pass as shown in~\eqref{eqn:matrix} and $p_{i_{\mathrm{avg}}}$, we estimate FER $\hat{p}_{i}$ experienced by the rate $r_i$ turbo encoded packets as $\hat{p}_{i} = \sum_{j} \beta_{ij} \, p_{j_{\mathrm{avg}}}$. Over a communication pass of duration $T$ seconds, total time spent transmitting at rate $r_i$ is $\beta_i \, T$, and the number of symbols sent at rate $r_i$ is $\hat{n}_i = (r_i \, R_b \, \beta_i \, T)/L$. Therefore, the proportion of symbols $\pi_i$ that are sent in rate $r_i$ turbo encoded packets over the communication pass is
\begin{equation}
\pi_i = \frac{ \hat{n}_i}{ \sum_j  \hat{n}_j} = \frac{ r_i \, \beta_i }{\sum_j  r_j \, \beta_j}.
\end{equation}
Assuming that the proportion holds true for each file transmission, simplified mean and variance of $\tilde{y}$ are given as $\mean{\tilde{y}} = \sum_{i} \pi_i \, {\paren{K_{\text{S}} + \Theta + y}} \, \hat{p}_{i}$ and $\operatorname{Var}\paren{\tilde{y}} = \sum_{i} \pi_i \, {\paren{K_{\text{S}} + \Theta + y}} \, \hat{p}_{i} \, (1 -\hat{p}_{i})$, respectively. The value of $y$ hence can be computed using $\mean{\tilde{y}}$, $\operatorname{Var}\paren{\tilde{y}}$, and~\eqref{eqn:y_value}.\\

\subsubsection{Adaptive DCSM}
\label{Sec:Adaptive_DCSM}
$\quad$\\*[-0.75em]

In this approach, $y_m$ is computed for each file separately at the spacecraft, i.e., $y_1, y_2, \cdots, y_{ F_{\mathrm{tx}} } $, on the basis of the proportion of symbols $\pi_i^{(m)}$ transmitted at a particular turbo code rates $r_i$ over the duration of the $m$-th file transfer and corresponding FERs. The actual FER experienced by each transmitted packet cannot be known at the spacecraft. However, corresponding to the use of each turbo code rate $r_i$, the maximum $p_{i_{\mathrm{max}}} = P_E(\text{SNR}_{i_{\mathrm{max}}})$ and minimum $p_{i_{\mathrm{min}}} = P_E(\text{SNR}_{i_{\mathrm{min}}})$ value of FER is a known quantity, and the average FER $p_{i_{\mathrm{avg}}} = \paren{p_{i_{\mathrm{max}}} + p_{i_{\mathrm{min}}}}/2$ can be easily computed. 

In \emph{adaptive best} DCSM, we use $p_{i_{\mathrm{min}}}$ as FER for each rate $r_i$ turbo encoded symbol transmission. Thus, resulting in \[ \mean{\tilde{y}_{m}} = \sum_{i} \pi_i^{(m)} \, {\paren{K_{\text{S}} + \Theta + y_{m}}} \, p_{i_{\mathrm{min}}}, \quad \mathrm{and}\] \[\operatorname{Var}\paren{\tilde{y}_{m}} = \sum_{i} \pi_i^{(m)} \, {\paren{K_{\text{S}} + \Theta + y_{m}}} \, p_{i_{\mathrm{min}}} \, (1 - p_{i_{\mathrm{min}}}).\]
Similarly, in \emph{adaptive worst} and \emph{adaptive average} DCSM, we use $p_{i_{\mathrm{max}}}$ and $p_{i_{\mathrm{avg}}}$, respectively, as FER for each rate $r_i$ turbo encoded symbol transmission. 

For all these cases, $y_{m}$ is computed for each transmitted file separately using its corresponding $\mean{\tilde{y}_{m}}$, $\operatorname{Var}\paren{\tilde{y}_{m}}$ and~\eqref{eqn:y_value}. The proportion  $\pi_i^{(m)}$ over the duration of $\paren{K_{\text{S}} + \Theta}$ symbols transmission from the $m$-th file is recorded and $y_m$ is computed at the end of $\paren{K_{\text{S}} + \Theta}$-th symbol transmission.

\subsection{Genie Method}


The premise of the \emph{genie} method is that the distant spacecraft has exact knowledge of actual channel bit-SNR condition and FERs corresponding to each transmitted turbo packets, and in turn, $\alpha_i$ is a known quantity. Keeping this important fact under consideration, the \emph{genie} approach is categorized into two types; a) \emph{genie I}, and b) \emph{genie II}. \\

\subsubsection{Genie I}
$\quad$\\*[-0.75em]

The features of \emph{genie I} are the same as that of the DCSM except that the actual future channel condition at the Earth station is known to the spacecraft {\it a priori}. Stepwise execution of \emph{genie I} is the same as that of the DCSM presented in Section III.A, except that in step ii), an appropriate turbo encoder is selected by the spacecraft itself based on the knowledge of future channel bit-SNR condition at the communicating Earth station.

The length, $L_p^{(r_i)}$, and data transmission rate, $R_d^{(r_i)}$, for code rate $r_i$ turbo encoded packets are given by~\eqref{eqn:pkt_length} and~\eqref{eqn:data_rate}, respectively. Over the duration of a communication pass, the number of symbols $ N_{\mathrm{tx}}^{(gI)} $ transmitted is 
\begin{equation}
\label{eqn:symbols_tx}
N_{\mathrm{tx}}^{(gI)} = \sum_{r_i \neq 0} \alpha_i \, r_i \, T  \, R_b \, \frac{1}{(K+36)}, 
\end{equation}
where the superscript ``$(gI)$'' is to represent that all this corresponds to the \emph{genie I} method. The number of failures, $\tilde{y}_m$, out of $\paren{K_{\text{S}} + \Theta + y_m}$ transmitted symbols of the $m$-th file is Gaussian distributed and is given as 
\begin{equation*}
\tilde{y}_m \sim \mathcal{N} \paren{\sum_{n=1}^{\paren{K_{\text{S}} + \Theta + y_m}} p_{i}^{(n)}, \sum_{n=1}^{\paren{K_{\text{S}} + \Theta + y_m}} p_{i}^{(n)} (1-p_{i}^{(n)})},
\label{eqn:y_genie}
\end{equation*}
where $p_i^{(n)}$ is the FER of $n$-th transmitted symbol of a file when optimal rate $r_i$ is used. Similar to that of the DCSM, average channel throughput $T_{H}^{(gI)}$, in terms of number of symbols received per second can be easily obtained. Based on how $y_m$ is calculated, {\em genie I} method is further divided into fixed and adaptive {\em genie I}.

In the fixed {\em genie I}, using statistics of the past communication pass, we know $\alpha_i$. Unlike in DCSM, FER $\hat{p}_{i}$ experienced by the rate $r_i$ turbo encoded packets are $\hat{p}_{i} = \sum_{j} \beta_{ij} \, p_{j_{\mathrm{avg}}} = p_{i_{\mathrm{avg}}} $. Over a communication pass of length $T$ seconds, the total time spent transmitting at rate $r_i$ is $T_i = \alpha_i T$ and the proportion of symbols $\pi_i$ transmitted in rate $r_i$ turbo encoded packets is
\begin{equation}
\pi_i = \frac{ \hat{n}_i}{ \sum_j  \hat{n}_j} = \frac{ r_i \, \alpha_i }{ \sum_j  r_j \, \alpha_j } = \frac{ r_i \, \alpha_i }{\sum_j  r_j \, \alpha_j}.
\end{equation}
Thus obtained $\mean{\tilde{y}}$ and $\operatorname{Var}\paren{\tilde{y}}$ is $\sum_{i} \pi_i \, {\paren{K_{\text{S}} + \Theta + y}} \, \hat{p}_{i}$ and $\sum_{i} \pi_i \, {\paren{K_{\text{S}} + \Theta + y}} \, \hat{p}_{i} \, (1 -\hat{p}_{i})$, respectively. 

For the adaptive {\em genie I}, the approach is same as that of adaptive DCSM described in Section~\ref{Sec:Adaptive_DCSM}. The details are the same except that the actual channel bit-SNR profile is used here instead of predicted channel bit-SNRs of DCSM.\\

\subsubsection{Genie II}
$\quad$\\*[-0.75em]

To evaluate performance of the \emph{genie} in the absence of RaptorQ codes, we also define the  \emph{genie II} method, characterized by a) dynamic selection of turbo codes at physical layer, and b) exact knowledge of future channel bit-SNRs. In the absence of RaptorQ codes at the application layer, \emph{genie II} lacks the capability of transmitting $y_m$ additional symbols of the $m$-th file and real-time feedback on the number of unsuccessful packets at the receiver. 
A stepwise elaboration on execution of the \emph{genie II} is given below:
\begin{enumerate}[label=\roman*)]
	\item At the spacecraft, each file to be telemetered is divided into ADUs of size $K = 8920$ bits and turbo encoded with the turbo encoder that maximizes instantaneous channel throughput at a given bit-SNR at the Earth station $\mathrm{RTT}/2$ into the future.
	\item Utilizing knowledge of FERs corresponding to the bit-SNR, one or
	multiple copies of each packet is transmitted so that its probability of success is approximately 1. 
	\item Once all the ADUs of current file is telemetered, transmission of ADUs from the following file begins.
\end{enumerate}

The length of a rate $r_i$ turbo encoded packet is given by~\eqref{eqn:pkt_length}. The data transmission rate, $R_d^{(r_i)}$, for code rate $r_i$ turbo code is $K\, R_b \, r_i/(K+36)$ bps. Over the duration of a communication pass, the number of symbols, $ N_{\mathrm{tx}}^{(gII)} $, transmitted is computed using~\eqref{eqn:symbols_tx}.

The number of unsuccessful packets out of $N_{\mathrm{tx}}^{(gII)}$ transmissions is approximately Gaussian distributed with mean $\sum_{n=1}^{N_{\mathrm{tx}}^{(gII)}} p_{i}^{(n)}$, and variance $\sum_{n=1}^{N_{\mathrm{tx}}^{(gII)}} p_{i}^{(n)} (1-p_{i}^{(n)})$. Utilizing this, average channel throughput $T_{H}^{(gII)}$, in terms of number of symbols successfully received at the receiver per second can be easily obtained.

\subsection{Static Method}


In \emph{static} method a fixed (8920, 1/2) turbo encoder is used throughout a communication pass. The \emph{static} method can be categorized into two types; a) \emph{static I}, and b) \emph{static II}. \\

\subsubsection{Static I}
$\quad$\\*[-0.75em]

Features of the \emph{static I} approach are a) RaptorQ at application layer, b) (8920, 1/2) turbo code at physical layer, c) $y_m$ additional symbols transmitted for the $m$-th file and d) real-time feedback on additional number of symbols to be transmitted from a file.
Stepwise execution of \emph{static I} is the same as that of the DCSM presented in Section III.A  except that in step ii) the (8920, 1/2) turbo encoder is always used.

The turbo encoded packet length $L_p$ and data transmission rate $R_d$ remains constant throughout a communication pass, and are given as 
\begin{align}
\label{eqn:static_pkt_length}
L_p &= 2 \, (36 + K)  \quad \mathrm{bits,}\\
R_d &= \frac{(K-32)}{2 \,(K+36)} \, R_b \quad \mathrm{bps}.
\end{align} 

The number of packets $ N_{\mathrm{tx}}^{(sI)} $ transmitted over the duration of a communication pass is given as
\begin{align}
N_{\mathrm{tx}}^{(sI)}  = T  \, R_b \, \frac{1}{2 \, (K+36)}.
\label{eqn:static_num_of_pkt}
\end{align}
The number of failed transmissions $\tilde{y}_m$ out of $\paren{K_{\text{S}} + \Theta + y_m}$ transmitted symbols of the $m$-th file is Gaussian distributed  
  
\begin{equation}
\label{eqn:y_static}
 \tilde{y}_m \sim \mathcal{N} \paren{\sum_{n=1}^{\paren{K_{\text{S}} + \Theta + y_m}} p_{1 \over 2}^{(n)}, \sum_{n=1}^{\paren{K_{\text{S}} + \Theta + y_m}} p_{1 \over 2}^{(n)} (1-p_{1 \over 2}^{(n)}) },
\end{equation}
where $p_{1 \over 2}^{(n)}$ is FER of the $n$-th transmitted symbol of the $m$-th file with $1/2$ rate turbo code. Static I method can further be divided into fixed method and adaptive method based on how $y_m$ is calculated.

In the fixed {\em static I} method, 
the use of $1/2$ rate turbo codes can be viewed as a scenario in which an arbitrary channel prediction model always predicts future channel bit-SNRs to be in Range E. 
Thus, the range prediction matrix $\boldsymbol\beta$ as shown in Table~\ref{table:statistics} has $\beta_{ij} = 1.0$ for $j = E$ and $0.0$ otherwise. 
Using the $\boldsymbol{\beta}$ and $p_{\mathrm{E}_{\mathrm{avg}}}$ (average FER of range $E$), an estimate of FER $\hat{p}_{\mathrm{E}}$ experienced by the rate $1/2$ turbo encoded packets is $\hat{p}_{\mathrm{E}} = p_{\mathrm{E}_{\mathrm{avg}}}$.
Similarly, $\pi_{\mathrm{E}} = 1$ and $\pi_{\mathrm{j}} = 0$ for $j \neq E$. The mean $\mean{\tilde{y}}$ and variance $\operatorname{Var}\paren{\tilde{y}}$ of $\tilde{y}$ is $\paren{ {\paren{K_{\text{S}} + \Theta + y}} \, \hat{p}_{\mathrm{E}}}$ and  $\paren{ {\paren{K_{\text{S}} + \Theta + y}} \, \hat{p}_{\mathrm{E}} \, (1 -\hat{p}_{\mathrm{E}})}$, respectively.

In adaptive {\em static I} method, we again compute $y$ for each file separately at the spacecraft, i.e., $y_1, y_2, \cdots, y_{ F_{\mathrm{tx}} }$ on the basis of the proportion of symbols $\pi_i^{(m)}$ transmitted at a particular turbo code rates $r_i$ over the duration of the $m$-th file transfer. Since, $\pi_{\mathrm{E}}^{(m)} = 1$ for all the files, the $y_{m}$ is same for all the files, that is, $y = y_1 = y_2 = \cdots = y_{ F_{\mathrm{tx}}}$. The maximum, minimum and average FER corresponding to the use of $1/2$ rate turbo code is $p_{\mathrm{E}_{\mathrm{max}}} = P_E(\text{SNR}_{E_{\mathrm{max}}})$, $p_{\mathrm{E}_{\mathrm{min}}}= P_E(\text{SNR}_{E_{\mathrm{min}}})$ and $p_{\mathrm{E}_{\mathrm{avg}}} = (p_{\mathrm{E}_{\mathrm{max}}} + p_{\mathrm{E}_{\mathrm{min}}})/2$, respectively. 

In adaptive best \emph{static I}, mean $\mean{\tilde{y}_m}$ and variance $\operatorname{Var}\paren{\tilde{y}_m}$ of $\tilde{y}_m$ is $\paren{ {\paren{K_{\text{S}} + \Theta + y_m}} \, p_{\mathrm{E}_{\mathrm{min}}}}$ and $\paren{ {\paren{K_{\text{S}} + \Theta + y_m}} \, p_{\mathrm{E}_{\mathrm{min}}} \, (1 - p_{\mathrm{E}_{\mathrm{min}}})}$, respectively. Similarly, the mean and the variance in case of adaptive worst \emph{static I} and adaptive average \emph{static I} is obtained by replacing $p_{\mathrm{E}_{\mathrm{min}}}$ by $p_{\mathrm{E}_{\mathrm{max}}}$ and $p_{\mathrm{E}_{\mathrm{avg}}}$, respectively.\\

\subsubsection{Static II}
$\quad$\\*[-0.75em]

To evaluate the performance of \emph{static} approach in the absence of RaptorQ codes, we define \emph{static II}, characterized by uninterrupted transmission of $(8920, 1/2)$ encoded turbo packets. Similar to \emph{genie II}, \emph{static II} lacks the capability of transmitting $y_m$ additional symbols of the $m$-th file and real-time feedback on the number of unsuccessful packets at the receiver. 

Each file to be telemetered is divided into $S_{\text{II}}$ ADUs of size $K = 8920$ bits. Each ADUs is individually turbo encoded and transmitted. Once all the ADUs from current the $m$-th file is sent, transmission of ADUs from the following $(m+1)$-th file begins. 

The data transmission rate $R_d$ is $R_d = K\, R_b/(2 \,(K+36)) $ bps. The number of failed transmissions $\tilde{y}_m$ out of $S_{\text{II}} $ transmitted symbols of the $m$-th file is also Gaussian distributed with mean $\mean{\tilde{y}_m} =  S_{\text{II}} \, \hat{p}_{\mathrm{E}}$ and variance $ \operatorname{Var}\paren{\tilde{y}_m} = S_{\text{II}} \, \hat{p}_{\mathrm{E}} \, (1 -\hat{p}_{\mathrm{E}})$. The number of packets $ N_{\mathrm{tx}}^{(sII)} $ transmitted during the communication pass is given in~\eqref{eqn:static_num_of_pkt}.
The number of unsuccessful packets out of $N_{\mathrm{tx}}^{(sII)}$ transmissions is approximately Gaussian distributed with mean $\sum_{n=1}^{N_{\mathrm{tx}}^{(sII)}} \hat{p}_{\mathrm{E}}$, and variance $\sum_{n=1}^{N_{\mathrm{tx}}^{(sII)}} \hat{p}_{\mathrm{E}} (1-\hat{p}_{\mathrm{E}})$. Utilizing this, average channel throughput $T_{H}^{(sII)}$, in terms of number of symbols successfully received at the receiver per second is obtained.

\section{Channel bit - SNR}
\label{sec:noise_temperature_bit_SNR}

	The distant spacecraft has a transmit antenna with gain $G_T$ that radiates a power $P_T$ in the direction of the DSN station's receiver antenna. On its way, the radiated power suffers from free space path loss and atmospheric attenuation. In practice, additional losses also occur due to losses in the transmitting and receiving equipment, de-pointing losses, and polarization mismatch losses. As we are evaluating baseline performance of the described methods, we assume that these additional losses are negligible. This section describes how the final channel bit-SNR is related to various noise factors. Some symbols used in this section are listed in Table III. 
	\begin{table}[h!]
		\begin{center}
			\caption{Table of symbols.}
			\begin{tabular}{l|p{14.2cm}}
				\toprule
				Symbol & Meaning\\ \toprule
				$\theta_e$ & antenna elevation angle \\
				$\theta_m$ & modulation index of carrier signal\\
				$D_{\text{wet}}^{(f)}$ & zenith wet path delay measured by an antenna operating at frequency $f$ \\ 
				$T_{\text{sky}}^{(f)}$ & zenith sky brightness temperature measured by an antenna operating at frequency $f$ \\
				$T_{\text{sky}}^{(f)}(\theta_e)$ & sky brightness temperature measured by an antenna operating at frequency $f$ and at an elevation angle of $\theta_e$ \\
				$L_{\text{atm}}^{(f)}$ & zenith atmospheric attenuation experienced by an antenna operating at frequency $f$\\
				$L_{\text{atm}}^{(f)} \paren{\theta_e}$ & atmospheric attenuation experienced by an antenna operating at frequency $f$ and at an elevation angle of $\theta_e$ \\
				$T_{\text{atm}}^{(f)}$ & zenith atmospheric noise temperature as seen by an antenna operating at frequency $f$ \\
				$T_{\text{atm}}^{(f)} \paren{\theta_e}$ & atmospheric noise temperature as seen by an antenna operating at frequency $f$ and at an elevation angle of $\theta_e$\\
				$T_{\text{op}}^{(f)}(\theta_e)$ & system operating noise temperature measured by an antenna operating at frequency $f$ and at an elevation angle of $\theta_e$ \\
				\bottomrule
			\end{tabular}
			\label{table:symbols}
		\end{center}
		\vspace*{-\baselineskip}
	\end{table}
	
	A receiving antenna operating at a frequency of $f$ and at an elevation angle $\theta_e$, with gain $G_R$, situated at a distance $D$ from the transmitting antenna receives power $P_R$ given as
	\begin{align}
	P_R &= P_T G_T  \paren{\frac{1}{L_{\text{FS}}^{(f)} \, L_{\text{atm}}^{(f)}(\theta_e)}} G_R, \notag \\
	&= P_T G_T  \paren{\frac{c}{4 \pi f D}}^2 \paren{\frac{1}{L_{\text{atm}}^{(f)}(\theta_e)}} G_R \quad \text{(Watt)},
	\label{eq:rcvd_power}
	\end{align}
	where $c= 3 \times 10^8$ m/s, $L_{\text{FS}}^{(f)}$ is free space path loss, and $L_{\text{atm}}^{(f)}(\theta_e)$ is \emph{atmospheric attenuation}.

\begin{figure}[h!]
		\centering
		\includegraphics[width=5.4in]{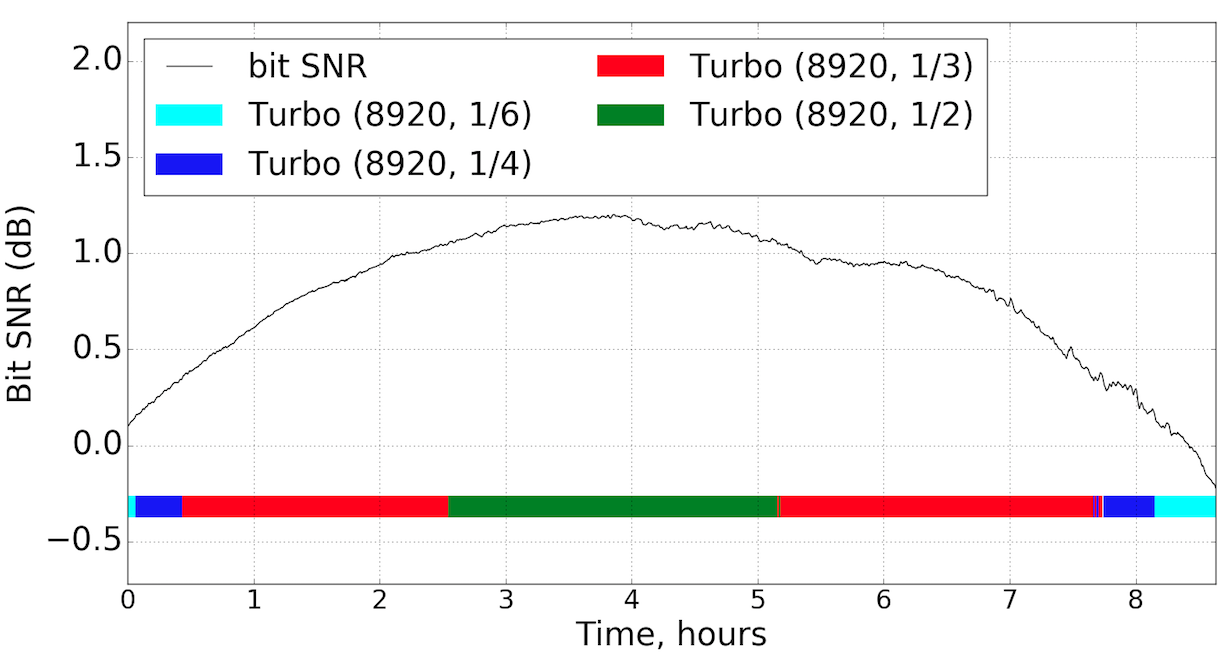}
		\caption{Channel bit-SNR (dB) vs time (hours) plot for DSS-25, along with turbo codes that maximize channel throughput at each bit-SNR during a communication pass.}
		\label{fig:Ka_eb}
		\vspace*{-\baselineskip}
	\end{figure}

	The DSN supports a wide range of telemetry modulation schemes \cite{Laboratory2000}. Modulation types used on MRO for Ka-band telemetry transmissions are binary phase shift keying (BPSK) on a square wave sub-carrier with the sub-carrier modulating the carrier, and BPSK directly on the carrier (no sub-carrier) \cite{JimTaylorSeptember2006}. With no sub-carrier or square wave sub-carrier, when a single telemetry channel is present and no ranging modulation is used, the telemetry channel data directly modulates the carrier with modulation index $\theta_m$ \cite{Laboratory2000}, \cite{Taylor2014}, \cite{JPL1983}. The $\theta_m$ is used to control the allocation of transmit power $P_T$ between carrier and data channels. The received carrier power $P_C$ and data power $P_D$, respectively, at the DSN station receiver antenna are given by \cite{Laboratory2000}
	
	\begin{eqnarray*}
		P_C = P_R \, \cos^2 \theta_m, \quad \mathrm{and} \quad
		P_D = P_R \, \sin^2 \theta_m,
		\label{eq:p_d}
	\end{eqnarray*}
	where $ P_R = P_C + P_D$. The input energy per bit $E_b$ to noise spectral density $N_0$ ratio (bit-SNR) of the communication channel as measured at DSN station is
	
	\begin{equation}
	\frac{E_b}{N_0} = \frac{P_D}{N_0 \, R_{b}} = \frac{P_R \, \sin^2 \theta_m}{N_0 \, R_{b}},
	\label{eq:bit_SNR}
	\end{equation}
	where $N_0 = K_b \, T_{\text{op}}^{(f)}(\theta_e)$ is one-sided noise spectral density referenced at the input to the receiver antenna's low-noise amplifier. The $K_b = 1.380622 \times 10^{-23}$ Watt/(Hz K) is the Boltzmann's constant and $T_{\text{op}}^{(f)}(\theta_e)$ is the system operating noise temperature. An example of channel bit-SNR plot of a communication pass for DSS-25 is presented in Fig.~\ref{fig:Ka_eb}. 
	
	\subsection{Atmospheric attenuation}
	\label{sec:Brightness_temperature_to_Atmospheric_attenuation}
	Presence of gaseous components and water (rain, clouds, snow and ice) in the troposphere and the ionosphere causes attenuation of waves as they propagate through the atmosphere. This attenuation experienced by a receiver antenna operating at a frequency $f$ at an elevation angle of $\theta_e$ is called {\em atmospheric attenuation} $L_{\text{atm}}^{(f)}(\theta_e)$ and is given as

	\begin{equation}
	L_{\text{atm}}^{(f)}(\theta_e) = \paren{\frac{T_p}{T_p -  T_{\text{atm}}^{(f)}}} \frac{1}{\sin \theta_e},
	\label{eq:atm_att}
	\end{equation}  
	where $T_p = 275 $ K is the physical temperature of the atmosphere, and $T_{\text{atm}}^{(f)}$ is {\em zenith atmospheric noise temperature} as seen by the antenna. 	
	
	\subsection{System operating noise temperature}
	\label{sec:attenuation_to_operating}
	
	The receiver {\em system operating noise temperature} $T_{\text{op}}^{(f)}(\theta_e)$ varies as a function of $\theta_e$ due to changes in the path length through the atmosphere and ground noise received by the side-lobe pattern of the antenna at a given frequency $f$ of operation. The $T_{\text{op}}^{(f)}(\theta_e)$ consists of an {\em antenna-microwave component} $T_{\text{AMW}}^{(f)}(\theta_e)$ and a {\em sky component} $T_{\text{sky}}^{(f)}(\theta_e)$ as shown is~\eqref{eq:top}. The $T_{\text{AMW}}^{(f)}(\theta_e)$ represents the contribution of the antenna and microwave hardware, and the $T_{\text{sky}}^{(f)}(\theta_e)$ represents the contribution of the atmospheric noise plus the cosmic microwave background noise. The $T_{\text{op}}^{(f)}(\theta_e)$ is given by
	
	\begin{align}
	T_{\text{op}}^{(f)}(\theta_e) &=   T_{\text{AMW}}^{(f)}(\theta_e) + T_{\text{sky}}^{(f)}(\theta_e) \notag \\
	&=  \left[ T_1 + T_2 e^{-a \theta_e} \right] + \left[  T_{\text{atm}}^{(f)} (\theta_e) + T^{'}_{\text{cmc}} (\theta_e) \right], 
	\label{eq:top}
	\end{align}
	where $T_1$, $T_2$, and $a$ are antenna-microwave noise temperature parameters, $T_{\text{atm}}^{(f)}(\theta_e)$ is the atmospheric noise temperature, and $T^{'}_{\text{cmc}}(\theta_e)$ is effective cosmic background noise at an elevation angle $\theta_e$. 
	The $T_{\text{atm}}^{(f)}(\theta_e)$ and $T^{'}_{\text{cmc}}(\theta_e)$ are given as 
	
	\begin{align}
	T_{\text{atm}}^{(f)}(\theta_e) &= T_M \left[ 1 - \frac{1}{L^{(f)}(\theta_e)}\right], \quad \text{K},\\
	T^{'}_{\text{cmc}} (\theta_e) &= \frac{T_{\text{cmc}}}{L^{(f)}(\theta_e)}, 
	\label{eq:T_cmc}
	\end{align}
	where $L^{(f)}(\theta_e) = 10^{\left( L_{\text{atm}}^{(f)}(\theta_e)/10\right) }$ is a dimensionless quantity with $L_{\text{atm}}^{(f)}(\theta_e)$ in dB, $T_M = 255 + 25 \times \text{CD}$ is the atmosphere mean effective radiating temperature (K), $0 \leq \text{CD} \leq 0.99$ is cumulative distribution, and $T_{\text{cmc}} = 2.725$ K is cosmic microwave background temperature.

	\subsection{Trace based analysis}
	\label{sec:trace_based} 
	
	Sky brightness temperatures \cite{Shambayati2008} have been measured for more than 20 years at Madrid deep space communication complex (DSCC), 17 years at Goldstone DSCC and 9 years at Canberra DSCC \cite{Laboratory2000}. For the trace based analysis and simulation, we use the recorded advanced water vapor radiometer (AWVR) measurement data of {\em zenith wet-path delay} $D_{\text{wet}}^{(31.4)}$ measurements for 31.4 GHz communication channel. Zenith wet path delay $D_{\text{wet}}^{(31.4)}$ along with time, antenna elevation, and antenna azimuth measurements taken every 30 s interval over the duration of multiple communication passes with Cassini at Goldstone DSCC for 31.4 GHz channel are archived in Planetary Data System (PDS). In order to use those data for conducting trace based analysis of Earth-spacecraft communication and to design a future channel bit-SNR prediction method, first we need every 1 second interval data. Since we have a sufficient number of samples of each communication pass to reconstruct the data, we use cubic basis-spline interpolation \cite{Schoenberg1988}, \cite{Unser1999} to fit those set of data points and generate a continuous signal representation.
	
	Additional delay incurred to a signal due to water content of the atmosphere is called wet path delay. The zenith sky brightness temperature $T_{\text{sky}}^{(f)}$ as seen from the ground is defined as the noise contribution of the entire atmosphere plus the attenuated noise contribution of the cosmic microwave background along the direction of zenith for an antenna operating at a frequency $f$ \cite{DavidMorabito2016}. In \cite{S.J.Keihm2001}, the authors present opacity vs wet path delay curves based on real measurement data over a period of one year at Goldstone site. Using those measurements for 31.4 GHz channel, the zenith wet path delay $D_{\text{wet}}^{(31.4)}$ is converted into water vapor opacity $\tau^{(31.4)}$ \cite{S.J.Keihm2001}. The \emph{zenith sky brightness temperature} experienced by the antenna operating at 31.4 GHz is computed from the $\tau^{(31.4)}$ \cite{inversion_algorithm} as
	
	\begin{equation}
	T_{\text{sky}}^{(31.4)} = 275 - 272 \, \exp(-\tau^{(31.4)}), \quad \mathrm{K}.
	\label{eqn:T_sky_31}
	\end{equation}
	Using $T_{\text{sky}}^{(31.4)}$, zenith atmospheric noise temperature $T_{\text{atm}}^{(31.4)}$ as seen by the antenna operating at $f = 31.4$ GHz is computed using the following relation \cite{Shambayati2008}:
	
	\begin{equation}
	T_{\text{atm}}^{(31.4)} = T_p \left( \frac{T_{\text{sky}}^{(31.4)} - T_{\text{cmc}}}{T_p - T_{\text{cmc}}} \right) \quad \mathrm{K}. 
	\label{eqn:T_atm_31}
	\end{equation}
	
	Since our focus is on Ka-band ($f = 32$ GHz) communication, we present a formulation applicable to that band here. The relationship between $T_{\text{atm}}^{(32)}$ and $T_{\text{atm}}^{(31.4)}$ is given as
	\begin{equation}
	T_{\text{atm}}^{(32)} = T_{\text{atm}}^{(31.4)} + 5 \left( 1 - \exp^{\left( -0.008 \, T_{\text{atm}}^{(31.4)}\right) } \right), \quad \mathrm{K}.
	\label{eqn:T_atm_32}
	\end{equation} 
	
	To summarize, using the traces of zenith wet path delay $D_{\text{wet}}^{(31.4)}$ and antenna elevation angle profile of a communication pass for an antenna operating at $31.4$ GHz, we first compute the value of $T_{\text{atm}}^{(32)}$ using~\eqref{eqn:T_sky_31} -~\eqref{eqn:T_atm_32}. Utilizing $T_{\text{atm}}^{(32)}$ and~\eqref{eq:atm_att}, the value of $L_{\text{atm}}^{(32)}(\theta_e)$ is obtained. Using the $L_{\text{atm}}^{(32)}(\theta_e)$ along with~\eqref{eq:top} -~\eqref{eq:T_cmc}, the value of $T_{\text{op}}^{(f)}(\theta_e)$ is obtained. Finally, using traces of $T_{\text{op}}^{(f)}(\theta_e)$, $L_{\text{atm}}^{(32)}(\theta_e)$,~\eqref{eq:rcvd_power} and~\eqref{eq:bit_SNR}, channel bit-SNR of a communication pass is obtained.

\section{Prediction Model}
\label{sec:prediction_model}
The fact that the weather condition in an area has a certain consistency during a small period of time allows the design of a useful channel prediction model and it is sufficient to predict future channel conditions once every second. All the measurable factors that characterize channel condition can be summarized into channel bit-SNR value. Hence, we develop a model that predicts channel bit-SNR values one RTT into the future. We proceed with the future channel bit-SNR prediction in two phases: preliminary  and real-time. 

\subsection{Preliminary prediction}
Channel bit-SNR, represented as $X_t$ at a particular instant $t$ during a communication pass depends on distance $D$ between the communicating spacecraft and the Earth station receiver antenna, receiver antenna elevation angle $\theta_e$, and randomness introduced due to weather effects. Let $t = 0, 1, \cdots, T_e$ be the time instant in seconds with $t = 0$ and $t = T_e$ representing the beginning and the end of a communication pass, respectively. We denote actual temporal channel bit-SNR of a communication pass by a time series $ \mathbf{X} =  \set{X_t | t = 0, 1, 2, \cdots, T_e} $. For each $X_t$, we will obtain an estimate $\hat{X}_t$. Thus, the obtained preliminary estimates of temporal channel bit-SNR $\mathbf{X}$ is denoted by another time series $ \mathbf{\hat{X}} =  \set{\hat{X}_t | t = 0, 1, 2, \cdots, T_e} $.

NASA's DSN supports multiple spacecrafts and service users. All the services and activities requiring DSN are scheduled in three phases: long range planning and forecasting (starts six months to one year before execution), mid-range scheduling (starts 4-5 months before execution), and near real-time scheduling (starts 8 weeks before the execution through the execution) \cite{MarkD.Johnston2012}. Thus, for each communication pass, we have information of the Earth and the Mars geometry, pass duration, visibility, and elevation angle of communicating DSN antenna in advance. Let us denote antenna elevation angle over a pass duration by a time series $\boldsymbol{\theta} =  \set{\theta_t | t = 0, 1, 2, \cdots, T_e}$, where $\theta_t$ represents the DSN antenna elevation angle at time $t$. Another time series $ \mathbf{D} =  \set{D_t | t = 0, 1, 2, \cdots, T_e} $ represents the Earth-Mars distance over the pass duration and $D_t$ is the Earth-Mars range at time $t$.

During communication with a distant spacecraft, the DSN site continuously records sky-brightness temperature measurements. Let us denote actual temporal sky brightness temperature of a communication pass as measured by the DSN site by a time series $ \mathbf{T_{\text{sky}}} =  \set{T_t^{(\text{sky})} | t = 0, 1, 2, \cdots, T_e} $, where $T_t^{(\text{sky})}$ is the actual sky-brightness temperature at time $t$. For each time index $t$, an average of sky-brightness temperature $\mathcal{T}_t^{(\text{sky})}$ over the duration of past `$N$' seconds, is computed 
\begin{equation}
\mathcal{T}_t^{(\text{sky})} = \sum_{i = 0}^{N-1} \frac{T_{t-i}^{(\text{sky})}}{N},
\label{eq:prel}
\end{equation}  
where $\mathcal{T}_t^{(\text{sky})}$ is the computed {\em a-priori} sky brightness temperature for the instant $t$ and $\mathbfcal{T}_t^{(\text{sky})} = \set{\mathcal{T}_t^{(\text{sky})} | t = 0, 1, 2, \cdots, T_e}$ represents the computed {\em a-priori} sky-brightness temperature over the pass duration. For our simulation purpose we are using $N = 10$, i.e, a duration of 10 seconds. Since immediate past few seconds of weather condition has higher correlation with present condition as compared to the conditions a few hours or days before, $N = 10$ gives a better approximation of $\mathcal{T}_t^{(\text{sky})}$.

For each time instant $t$, using $\mathcal{T}_t^{(\text{sky})}$ along with $\theta_{t + \mathrm{RTT} }$ and $R_{t + \mathrm{RTT} }$, we compute the corresponding value of channel bit-SNR $\hat{X}_{t + \mathrm{RTT} }$. The obtained $\hat{X}_{t + \mathrm{RTT} }$ is a preliminary estimate of actual channel bit-SNR $X_{t + \mathrm{RTT}}$. We use the obtained temporal channel bit-SNR estimate $\mathbf{\hat{X}}$ as prior channel bit-SNR information during real time prediction phase.

\subsection{Real-time prediction}
In order to further enhance the accuracy of preliminary prediction, we utilize the real-time measurements and the concept of an AR(1) process.\\ 

\subsubsection{AR(1) process}
$\quad$\\*[-0.75em]

By analyzing autocorrelation and partial autocorrelation (PAC) \cite{PeterJ.Brockwell2002}, \cite{GeorgeEPBox2008} of $\mathbf{X}$ of different communication passes, we first verified that the bit-SNR time series $ \textbf{X} =  \set{X_t | t = 0, 1, 2, \cdots, T_e} $ can be represented by an AR(1) process. For any time series $\textbf{X}$, the PAC at lag $k$ is the autocorrelation between $X_{t}$ and $X_{{t-k}}$ with the linear dependence of $X_{t}$ on $X_{{t-1}}$ through $X_{{t-k+1}}$ removed. 

The partial autocorrelation function (PACF) of $\textbf{X}$ at lag $k \in \set{0, 1, 2,\cdots}$, denoted by $\alpha_{\mathbf{X}}(k)$, is very useful in identifying an auto-regressive (AR) process. A series can be represented as a pure AR process of order $p$, if i) the autocorrelation function dies out in an exponential or sinusoidal fashion, and ii) the partial autocorrelation cuts off after lag $p$. That is, if our original process is auto-regressive of order $p$ represented as AR($p$), then for $k > p$, we should have $\alpha_{\mathbf{X}}(k) = 0$. 

From the analysis of autocorrelation and PACF of channel bit-SNR of communication passes, it is observed that both the above mentioned conditions are satisfied. The PACF of the channel bit-SNR with lag of one second is 1, i.e., $\alpha_{\mathbf{X}}(1) = 1$, and is 0 with lag more than 1 second, i.e., $\alpha_{\mathbf{X}}(k) \simeq 0$ for $k > 1$. This implies that the channel bit-SNR $X_t$ depends on $X_{t-1}$ but not on previous values and hence can be defined by an AR(1) process given as 
\begin{equation}
X_{t}= c + \varphi \, X_{{t-1}}+ \varepsilon_{t}\,,
\end{equation} 
where $c$ is a constant, $\varphi \in (0,1)$  is a constant multiplicative factor, and $\varepsilon_{t}$ is a zero mean and constant variance white noise process at time $t$. 

The AR(1) process is a discrete time analogy of the continuous \emph{Ornstein-Uhlenbeck} process and can be cast into Ornstein-Uhlenbeck equivalent form \cite{Mikosch1998}. The Ornstein-Uhlenbeck process \cite{Uhlenbeck1930} is a stochastic Gauss-Markov process that describes the velocity of a massive Brownian particle under the influence of friction \cite{Bibbona2008}. Utilizing the Ornstein-Uhlenbeck equivalent form of AR(1) process, we can approximate the relationship between $X_t$ and $X_{t+n}$ as given below
\begin{equation}
\label{eq:mean_OU}
\operatorname {E} \left[ X_{{t+n}}|X_{t} \right] =\mu \left[1-\paren{1-\phi} ^{n} \right] + X_{t}\paren{1-\phi }^{n}, 
\end{equation}
where $X_{t+n}$ represents bit-SNR value $n$ periods into the future, and $\mu$ and $|\phi|<1\,$ are the model parameters. The $\mu = c / (1 - \varphi)$ is the long term mean and the $\phi = (1 - \varphi)$ is the rate of mean reversion of the process. Using equation~\eqref{eq:mean_OU}, based on current state of the time series we can forecast its value an arbitrary number of periods into the future.\\ 

\subsubsection{Prediction algorithm}
$\quad$\\*[-0.75em]

 At the communicating DSN station, for each time index $t$ during the period of communication, we compute the estimation error $e_t = X_t - \hat{X}_t$ between the actual measured bit-SNR $X_t$ and its preliminary estimation $\hat{X}_t$. The $e_t$ can also be characterized by an AR(1) process and hence the relation between $e_t$ and $e_{t+n}$ is given as
  \begin{equation}
 	\label{eq:mean_E}
 	\operatorname {E} \left[ e_{{t+n}}|e_{t} \right] =\mu_e \left[1-\paren{1-\phi_e} ^{n} \right] + e_{t}\paren{1-\phi_e }^{n},
 \end{equation}
where $e_{{t+n}}$ represents the estimation error $n$ periods into the future, $\mu_e$ is the long term mean and $|\phi_e|<1\,$ is the rate of mean reversion of the process.

Utilizing the value of $e_t$ and~\eqref{eq:mean_E}, we obtain an error estimate one second later $\hat{e}_{t+1}$ and one RTT into the future $\hat{e}_{t+\mathrm{RTT}}$. We then correct the preliminary prediction $\hat{X}_{t+1}$ by $\hat{e}_{t+1}$ amount so that the updated value is $\hat{X}_{t+1} = \hat{X}_{t+1} + \hat{e}_{t+1}$. Similarly, we update the preliminary prediction $\hat{X}_{t+\mathrm{RTT}}$ by $\hat{e}_{t+\mathrm{RTT}}$ amount so that the updated $\hat{X}_{t+\mathrm{RTT}}^{(\text{pred})} = \hat{X}_{t+\mathrm{RTT}} + \hat{e}_{t+\mathrm{RTT}}$. All the updated $\hat{X}_{t+\mathrm{RTT}}^{(\text{pred})}$ are considered as our final prediction and is represented in time series form as $ \mathbf{\hat{X}}_{\text{pred}} =  \set{\hat{X}_t^{(\text{pred})} | t = 0, 1, 2, \cdots, T_e} $. The process is presented in Algorithm~\ref{alg:prediction}.

\begin{algorithm}[h!]
	\caption{Channel bit-SNR prediction algorithm.}
	\KwData{$X_0$, $\mathbf{\hat{X}}$}
	initialization $t = 0$, $\mathbf{\hat{X}}_{\text{pred}} = \textbf{0}$\;
	\While{$t \leq T_e$}{
		read $X_t, \hat{X}_t$\;
		initialize $ e_t = X_t - \hat{X}_t$\;		
		predict $\hat{e}_{t+1}$, $\hat{e}_{t + \mathrm{RTT}}$\;
		update $\hat{X}_{t+1} = \hat{X}_{t+1} + \hat{e}_{t+1}$\;
		update $\hat{X}_{t+\mathrm{RTT}} = \hat{X}_{t+\mathrm{RTT}} + \hat{e}_{t+\mathrm{RTT}}$\;
		save $\hat{X}_{t+\mathrm{RTT}}^{\text{(pred)}} = \hat{X}_{t+\mathrm{RTT}}$
	}
	\label{alg:prediction}
\end{algorithm}

\begin{figure}[h!]
	\begin{center}
		\includegraphics[width=5.4in]{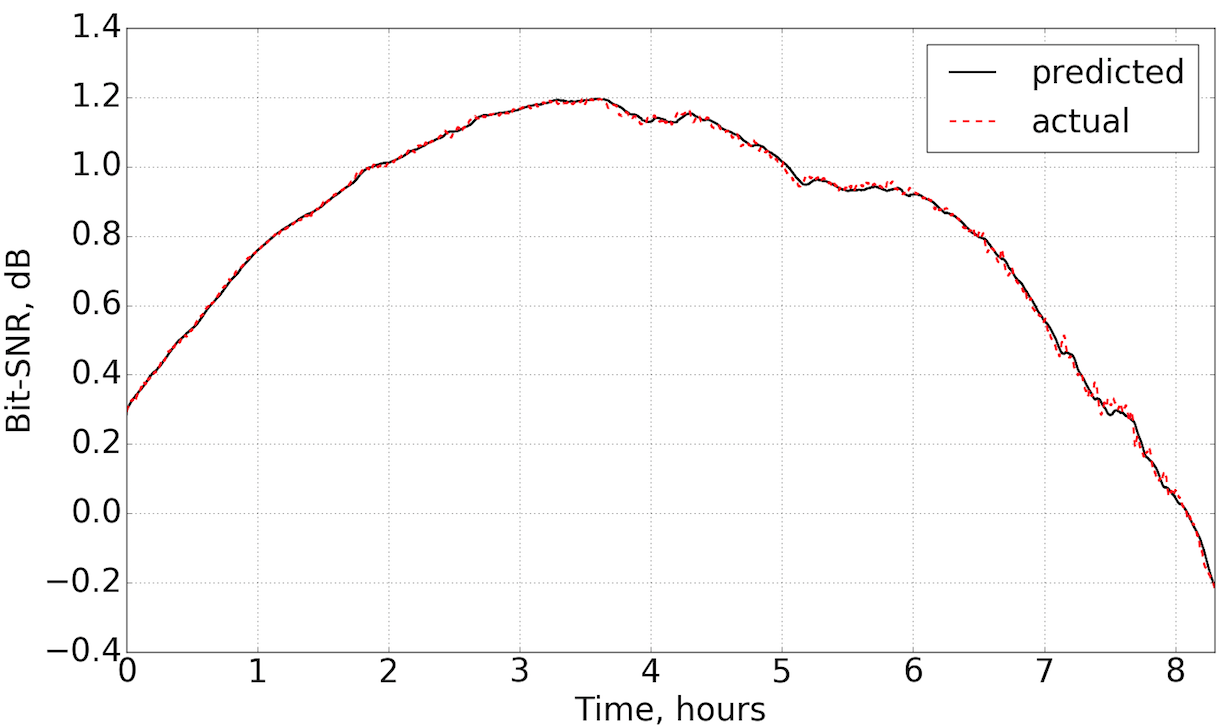}
		\caption{Predicted and actual temporal bit-SNR of a communication pass on September 26, 2005.}
		\label{fig:O_Sept}
	\end{center}
	\vspace*{-\baselineskip}
\end{figure}

One example of the predicted channel bit-SNR curve along with the actual bit-SNR curve of a communication pass obtained using Algorithm~\ref{alg:prediction} is shown in Fig.~\ref{fig:O_Sept}. We can see that our prediction model is capable of tracking the actual channel condition and predicting the rapid fluctuation efficiently.

\begin{table}[h!]
	\centering
		\caption{Channel bit-SNR estimation accuracy (in \%) during a communication pass on December 15, 2001 data.}
		\begin{tabular}{l|l|l|l|l|l|l}
			\toprule
			\multirow{2}{*}{\raisebox{-\heavyrulewidth}{\% of}} & \multirow{2}{*}{\raisebox{-\heavyrulewidth}{Ranges}} & \multicolumn{5}{ c }{Predicted SNR range} \\
			\cmidrule{3-7} 
			SNRs &  & Range A & Range B & Range C & Range D & Range E \\
			\midrule	
			7.196 & Range A & \cellcolor{gray!45}98.909 & 1.091 & 0.000 & 0.000 & 0.000\\
			4.907 & Range B & 1.417 & \cellcolor{gray!45}97.577 & 1.006 & 0.000 & 0.000\\
			3.749 & Range C & 0.000 & 0.000 & \cellcolor{gray!45}100.000 & 0.000 & 0.000\\
			15.751 & Range D & 0.000 & 0.000 & 0.826 & \cellcolor{gray!45}98.589 & 0.584\\
			68.397 & Range E & 0.000 & 0.000 & 0.000 & 0.128 & \cellcolor{gray!45}99.872\\
			\midrule
			\multicolumn{2}{ c |}{\% of predicted SNR} & 7.187 & 4.867 & 3.929 & 15.616 & 68.402\\
			\midrule
			\multicolumn{7}{ c }{\cellcolor{gray!45} Percentage of correct estimation: 99.493}\\
			\bottomrule
		\end{tabular}
		\label{table:one}
			\vspace*{-\baselineskip}
\end{table}

Table~\ref{table:one} demonstrates the percentages of estimation accuracy in individual ranges and the overall estimation accuracy during a communication pass of December 15, 2001. 7.196\% of channel bit-SNR is observed in range A, 4.907\% in range B, 3.749\% in range C, 15.751\% in range D, and 68.397\% in range E. For the case when channel bit-SNR was actually in range A, 98.909\% of time we predicted it to be in range A and 1.091\% of time in range B. Overall, we were able to correctly predict channel bit-SNR 99.493\% of time during the communication pass.

\begin{figure}[h!]
		\centering
		\includegraphics[width=0.9\columnwidth]{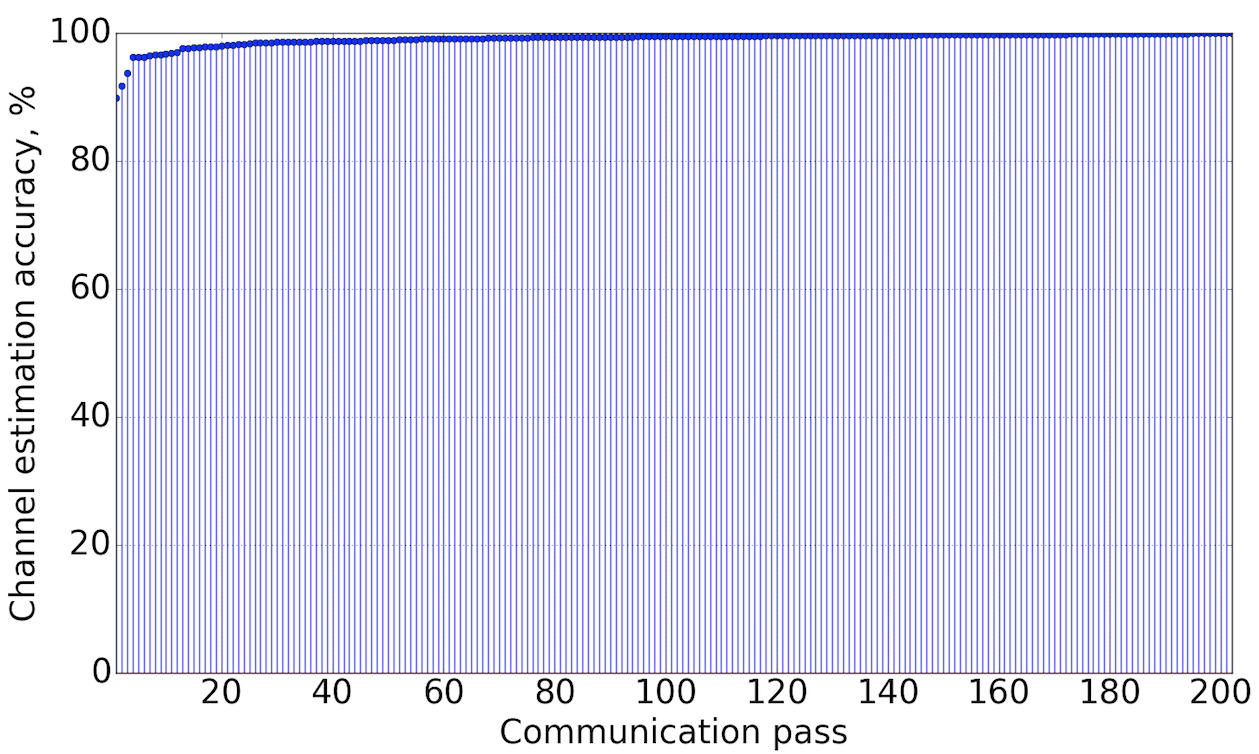}
		\caption{ Percentage of correct channel bit-SNR estimation plots obtained using our channel bit-SNR prediction model for passes under consideration. }
		\label{fig:estimation}
	 	\vspace*{-\baselineskip}
\end{figure}

Using the AWVR data of 202 different communication passes between year 2001 and 2011, we computed the percentage of correct estimation for each communication pass. The channel prediction algorithm is run once for each communication pass. For different communication passes, different percentages of correct estimation are observed, with maximum and minimum percentage of correct estimation being 100\% and 89.83\% respectively. Percentage of correct estimation achieved in each communication pass are arranged in increasing order and plotted against the communication pass as shown in Fig.~\ref{fig:estimation}. Among these 202 communication passes, our prediction accuracy is less than 90\% for only one pass. For the rest of the 201 passes, we have prediction accuracy more than 90\%. 

\section{Simulation Setup and Results}
\label{sec:simulation}
\subsection{Simulation setup and communication data rates}
MRO supports both the X-band and Ka-band communications. We are considering only the Ka-band communication scenario for our simulation purpose and are simulating data transmission between the MRO's high gain antenna (HGA) and 34-m beam-waveguide (BWG) antenna DSS-25 at Goldstone DSCC. In MRO, BPSK modulation is used for Ka-band transmission, and it has a limited number of channel modulation rates and modulation index values. The parameters used in our simulations are summarized in Table~\ref{table:data}. 
	
	\begin{table}[h!]
		\vspace*{-\baselineskip}
		\centering
			\caption{Parameters used in our simulation setup.}
			\begin{tabular}{l|l}
				\toprule
				Parameters & Values \\ \toprule
				Transmission power (MRO's HGA) & 34 Watt  \\ 
				Transmit antenna gain (MRO's HGA) & 56.4 dBi \\
				Receive antenna gain (DSS-25) & 79 dBi \\
				Modulation index (MRO's HGA) & 21.09375 \\
				Channel modulation rate (MRO's HGA) & 3 Msps\\
				Earth-MRO distance & $18 \times 10^{10}$ m\\
				File size & 50 MB \\
				$\Theta$ (RaptorQ codes) & 5 \\
				\bottomrule
			\end{tabular}
			\label{table:data}
		\vspace*{-\baselineskip}
	\end{table}

For simplicity, dynamic distance (and hence RTT) is not considered in this paper and the RTT between Earth-MRO is set to 20 minutes. For $K = 8920$ bits, Table~\ref{table:data_rates} gives the data rates $R_d$ (Mbps) for different simulation setups with different turbo code rates.

\addtolength{\tabcolsep}{-1pt} 
\begin{table}[h!]
	\centering
	\caption{ Data rates $R_d$ (Mbps) for $R_b = 3$  Mbps and turbo code rates for $K = 8920$ bits.}
	\label{table:data_rates}
	\begin{tabular}{c|cccc|cccc}
		\toprule
		\multirow{2}{*}{\raisebox{-\heavyrulewidth}{$R_b$}} & \multicolumn{4}{c|}{DCSM, \emph{Genie I} and \emph{Static I}} & \multicolumn{4}{c}{\emph{Genie II} and \emph{Static II}}\\
		\cmidrule{2-9}
		(Mbps) & $1/6$ &$1/4$ &$1/3$ & $1/2$ & $1/6$ &$1/4$ &$1/3$ & $1/2$\\ \toprule
		3 & 0.496 & 0.744 & 0.992 & 1.489 & 0.498 & 0.747 & 0.996 & 1.494 \\
		\bottomrule
	\end{tabular}
\end{table}
\addtolength{\tabcolsep}{1pt}

The simulation is run once for each communication pass for all the above mentioned methods. Each file is considered of fixed size 50 MB, resulting in $K_{\text{S}} = 45005$ with $L = 8888$ bits for DCSM, \emph{genie I}, and \emph{static I} without CRC. In case of \emph{genie II} and \emph{static II}, each file is divided into 44844 ADUs of length $K = 8920$ bits. In the simulation setup, we do not consider any system losses and hence the results are baseline performance achieved with the proposed mechanism.  

 For each communication pass, the total bits transmitted ($B_{\mathrm{Tx}}$) and the total data transmitted ($D_{\mathrm{Tx}}$) by the MRO (in Gbs), and the total bits received ($B_{\mathrm{Rcvd}}$) and the total data received ($D_{\mathrm{Rcvd}}$) by DSS-25 (in Gbs) over the duration of the pass is recorded. The $D_{\mathrm{Tx}}$ is a measure of actual file content transmitted, excluding all the headers and trailers. The total $D_{\mathrm{Tx}}$ and $D_{\mathrm{Rcvd}}$ over the duration of a communication pass depends on a number of factors, namely, duration of the communication pass, antenna elevation angle, channel bit-SNRs, weather dropout conditions, and the accuracy with which channel bit-SNRs are predicted. 
 
By dividing $D_{\mathrm{Tx}}$ and $D_{\mathrm{Rcvd}}$ by file size (50 MB = 0.4 Gbs), we use $F_{\mathrm{TxEqv}} = D_{\mathrm{Tx}} / 0.4$ and $F_{\mathrm{RcvdEqv}} = D_{\mathrm{Rcvd}} / 0.4$ to represent the equivalent number of files transmitted by the MRO and received by the DSS-25, respectively. $F_{\mathrm{Tx}}$ represents actual number of files transmitted and $F_{\mathrm{RcvdSuccess}}$ represents number of complete files that are successfully received over the duration of a communication pass.

\begin{figure}[h!]
		\centering
		\includegraphics[width=\columnwidth]{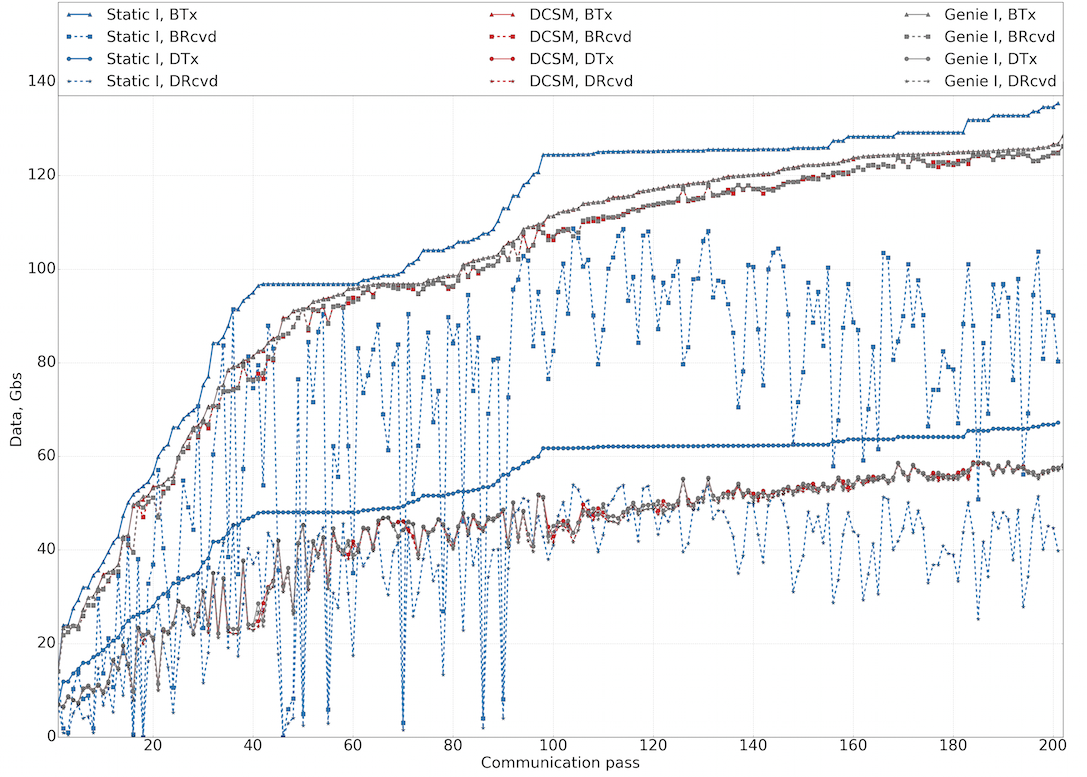}
	\caption{Bits transmitted ($B_{\mathrm{Tx}}$), bits received ($B_{\mathrm{Rcvd}}$), data transmitted ($D_{\mathrm{Tx}}$), and data received ($D_{\mathrm{Rcvd}}$) during each communication pass with adaptive  best approach.}
			\label{fig:adaptive_best}
\vspace*{-\baselineskip}
\end{figure}

Plots showing $B_{\mathrm{Tx}}$, $D_{\mathrm{Tx}}$, $B_{\mathrm{Rcvd}}$, and $D_{\mathrm{Rcvd}}$ during each communication pass with adaptive best approach are shown in Fig.~\ref{fig:adaptive_best}. For the clarity of presentation, the communication passes are arranged in increasing order in terms of the $B_{\mathrm{Tx}}$ by the MRO in \emph{static I} adaptive approach. As an example, on the 120$^{\text{th}}$ communication pass, $B_{\mathrm{Tx}}$ = 117.08 Gbs and $D_{\mathrm{Tx}}$ = 49.76 Gbs, and $B_{\mathrm{Rcvd}}$ = 113.67 Gbs and $D_{\mathrm{Rcvd}}$ = 48.87 Gbs, over the duration of the communication pass using the adaptive best DCSM.

\subsection{Intra-method performance evaluation}
By averaging over the 202 communication passes for each method, a summary of data (in Gbs) and file (count) transmissions during each communication pass (on average) is obtained. A summary of \emph{genie} approach is presented in Table~\ref{table:tx_data_genie}. With the \emph{genie II}, on average, $D_{\mathrm{Tx}} = 43.533 $ Gbs, and $ D_{\mathrm{Rcvd}} = 42.873$ Gbs, which is equivalent to $F_{\mathrm{TxEqv}} = 108.833$ and $F_{\mathrm{RcvdEqv}} = 107.183$ files in size, respectively. However, the actual number of files transmitted and successfully received, on average, are $F_{\mathrm{Tx}} = 108.446$ and $F_{\mathrm{RcvdSuccess}} = 62.673$, respectively. That is, the 42.873 Gbs of received data resulted in 62.673 complete file reception.

With the \emph{genie I}, on average, the $B_{\mathrm{Tx}}$, $D_{\mathrm{Tx}}$, $B_{\mathrm{Rcvd}}$, and $D_{\mathrm{Rcvd}}$ obtained with all the sub-classes are almost the same. However, there is huge difference in performance in terms of $F_{\mathrm{Tx}}$ and $F_{\mathrm{RcvdSuccess}}$. For almost the same $B_{\mathrm{Tx}}$, $D_{\mathrm{Tx}}$, $B_{\mathrm{Rcvd}}$, and $D_{\mathrm{Rcvd}}$ with all the \emph{genie I} sub-classes, there is remarkable difference in the number of files transmitted and successfully received. This difference comes from the value of $y$ used for each of the sub-classes. Looking at the $F_{\mathrm{RcvdSuccess}}$ values, we can observe that the fixed \emph{genie I} approach results in larger number of $F_{\mathrm{RcvdSuccess}}$ than that obtained with the adaptive \emph{genie I} approach. Since our primary objective is to deliver the maximum number of files successfully over the same duration of a communication pass, it can be concluded that the fixed \emph{genie I} is superior to the adaptive \emph{genie I}.

\addtolength{\tabcolsep}{-3pt}    
\begin{table}[h!]
	\vspace*{-\baselineskip}
	\centering
	\caption{ Summary of data and file transmissions on average during a communication pass with \emph{genie} approach.}
	\vspace*{-\baselineskip}
	\label{table:tx_data_genie}
	\begin{subtable}{0.5\textwidth}
		\caption{Summary of data (in Gbs) and their ratios.}
		\begin{tabular}{l|ccc|ccc|c}
			\toprule
			\multirow{4}{*}{\raisebox{-\heavyrulewidth}{Data}} & \multicolumn{6}{c|}{\emph{Genie I}} & \multirow{4}{*}{\raisebox{-\heavyrulewidth}{\emph{Genie II}}} \\
			\cmidrule{2-7}
			& \multicolumn{3}{c}{ Fixed $y$} & \multicolumn{3}{|c|}{ Adaptive $y$} & \\
			\cmidrule{2-4} \cmidrule{5-7}
			& $y$ & $2 y$ & $y_{\mathrm{min}}$ & Best & Average & Worst & \\ \toprule
			$B_{\mathrm{Tx}}$ & 100.565 & 100.565 & \cellcolor{gray!30}100.565 & 100.565 & 100.565 & 100.565 & 100.565 \\
			$D_{\mathrm{Tx}}$ & 43.377  & 43.377 & \cellcolor{gray!30}43.377 & 43.377 & 43.377 & 43.377 & 43.533 \\ 
			$B_{\mathrm{Rcvd}}$ & 98.188 & 98.187 & \cellcolor{gray!30}98.187 & 98.187 & 98.188 & 98.188 & 98.187 \\
			$D_{\mathrm{Rcvd}}$ & 42.720 & 42.720 & \cellcolor{gray!30}42.720 & 42.720 & 42.720 & 42.720 & 42.873 \\
			\toprule
			$D_{\mathrm{Tx}}$/$B_{\mathrm{Tx}}$ & 0.4313 & 0.4313 & \cellcolor{gray!30}0.4313 & 0.4313 & 0.4313 & 0.4313 & 0.4329 \\
			\rowcolor{gray!45}$D_{\mathrm{Rcvd}}$/$B_{\mathrm{Tx}}$ & 0.4248 & 0.4248 & \cellcolor{gray!30}0.4248 & 0.4248 & 0.4248 & 0.4248 & 0.4263 \\
			\rowcolor{gray!45}$D_{\mathrm{Rcvd}}$/$D_{\mathrm{Tx}}$ & 0.9848 & 0.9848 & \cellcolor{gray!30}0.9848 & 0.9848 & 0.9848 & 0.9848 & 0.9848 \\	
			$D_{\mathrm{Rcvd}}$/$B_{\mathrm{Rcvd}}$ & 0.4351 & 0.4351 & \cellcolor{gray!30}0.4351 & 0.4351 & 0.4351 & 0.4351 & 0.4366 \\ 			
			\bottomrule
		\end{tabular}
		\label{table:tx_data_genie_data}
	\end{subtable}
		
	\begin{subtable}{0.5\textwidth}
		\caption{Summary of the number of files.}
		\begin{tabular}{l|ccc|ccc|c}
			\toprule
			\multirow{4}{*}{\raisebox{-\heavyrulewidth}{Files}} & \multicolumn{6}{c|}{\emph{Genie I}} & \multirow{4}{*}{\raisebox{-\heavyrulewidth}{\emph{Genie II}}} \\
			\cmidrule{2-7}
			& \multicolumn{3}{c}{ Fixed $y$} & \multicolumn{3}{|c|}{ Adaptive $y$} & \\
			\cmidrule{2-4} \cmidrule{5-7}
			& $y$ & $2 y$ & $y_{\mathrm{min}}$ & Best & Average & Worst & \\ \toprule	
			$F_{\mathrm{TxEqv}}$ & 108.443 & 108.443 & \cellcolor{gray!30}108.443 & 108.443 & 108.443 & 108.443 & 108.833 \\
			$F_{\mathrm{RcvdEqv}}$ & 106.799  & 106.799 & \cellcolor{gray!30}106.799 & 106.799 & 106.800 & 106.799 & 107.183 \\
			\toprule
			\rowcolor{gray!45}$F_{\mathrm{Tx}}$ & 102.963 & 98.950 & \cellcolor{gray!30}106.176 & 108.023 & 105.886 & 81.367 & 108.446\\
			\rowcolor{gray!45}$F_{\mathrm{RcvdSuccess}}$ & 100.634 & 97.608 & \cellcolor{gray!30}102.689 & 77.307 & 102.314 & 81.245 & 62.673\\
			\bottomrule
		\end{tabular}
		\label{table:tx_data_genie_files}
		\vspace*{-\baselineskip}
	\end{subtable}
\end{table}
\addtolength{\tabcolsep}{1pt} 

As seen in the Table~\ref{table:tx_data_genie}, the fixed \emph{genie I} has three different sub-classes given as
\begin{enumerate}
	\item $y$ additional symbols with each file, calculation of $y$ is given in section III. 
	\item $2y$ additional symbols with each file, and
	\item $y_{\textrm{min}}$ additional symbols with each file. Minimum value $y_{\textrm{min}} = 453$ is obtained for a pass when bit-SNR of the entire pass duration is in Range E, and is predicted with 100\% accuracy.
\end{enumerate}

Among these sub-classes, fixed \emph{genie I} with $y_{\textrm{min}}$ results in largest number of $F_{\mathrm{RcvdSuccess}}$. Hence, it can be concluded that fixed \emph{genie I} with $y_{\textrm{min}}$ is superior to other \emph{genie I} subclasses, which in turn is superior to \emph{genie II} as well. 

The same pattern has been observed for \emph{static} and DCSM approaches as well. That is, fixed DCSM with $y_{\textrm{min}}$ is superior to other DCSM sub-classes, and fixed \emph{static I} with $y_{\textrm{min}}$ is superior to other \emph{static} sub-classes. For all three methods, we observed that fixed approach with $y_{\textrm{min}}$ is the best option to be considered (within each method) to achieve highest number of successful file deliveries. 

Complexity of implementation of fixed approach with $y_{\textrm{min}}$ is drastically low (in comparison to the other sub-classes), as we can always use the same $y = 453$ for all the files of all the communication passes. For each file, we are adding 1.00644\% of additional symbols, i.e., $ 1.00644 \% $ of 45010 is 453, by transmitting $y = 453$ additional symbols, which is negligible given the performance improvement that is achieved.

\subsection{Inter-method performance evaluation}

\begin{table}[h!]
	\vspace*{-\baselineskip}
	\centering
	\caption{ A comparative summary of data and file transmissions on average during a communication pass with DCSM, \emph{genie} and \emph{static}.}
	\vspace*{-\baselineskip}
	\label{table:tx_data_all}
	\begin{subtable}{0.5\textwidth}
		\caption{Summary of data (in Gbs) and their ratios.}
		\begin{tabular}{l|ccc|cc}
			\toprule
			Data & DCSM & \emph{Genie I} & \emph{Static I} & \emph{Genie II} & \emph{Static II}  \\
			& Fixed $y_{\mathrm{min}}$ & Fixed $y_{\mathrm{min}}$ & Fixed $y_{\mathrm{min}}$ &  &   \\ \toprule
			$B_{\mathrm{Tx}}$ & 100.569 & 100.565 & 106.769 & 100.565 & 106.181 \\
			$D_{\mathrm{Tx}}$ & 43.377  & 43.377 & 52.979 & 43.533 & 52.877\\ 
			$B_{\mathrm{Rcvd}}$ & 98.131 & 98.187 & 71.625 & 98.187 & 70.890\\
			$D_{\mathrm{Rcvd}}$ & 42.696 & 42.720 & 35.541 & 42.873 & 35.302\\
			\toprule
			$D_{\mathrm{Tx}}$/$B_{\mathrm{Tx}}$ & 0.4313 & 0.4313 & 0.4962 & 0.4329 & 0.4980\\
			\rowcolor{gray!45}$D_{\mathrm{Rcvd}}$/$B_{\mathrm{Tx}}$ & 0.4248 & 0.4245 & 0.3329 & 0.4263 & 0.3325\\
			\rowcolor{gray!45}$D_{\mathrm{Rcvd}}$/$D_{\mathrm{Tx}}$ & 0.9848 & 0.9843 & 0.671 & 0.9848 & 0.6676\\	
			$D_{\mathrm{Rcvd}}$/$B_{\mathrm{Rcvd}}$ & 0.4351 & 0.4351 & 0.4962 & 0.4366 & 0.4980 \\ 			
			\bottomrule
		\end{tabular}
		\label{table:tx_data_all_data}
	\end{subtable}
	
	\begin{subtable}{0.5\textwidth}
		\caption{Summary of the number of files.}
		\begin{tabular}{l|ccc|cc}
			\toprule
			Data & DCSM & \emph{Genie I} & \emph{Static I} & \emph{Genie II} & \emph{Static II}  \\
			& Fixed $y_{\mathrm{min}}$ & Fixed $y_{\mathrm{min}}$ & Fixed $y_{\mathrm{min}}$ &  &   \\ \toprule			
			$F_{\mathrm{TxEqv}}$ & 108.443 & 108.443 & 132.449 & 108.833 & 132.192\\
			$F_{\mathrm{RcvdEqv}}$ & 106.741 & 106.799 & 88.852 & 107.183  & 88.256\\
			\toprule
			\rowcolor{gray!45}$F_{\mathrm{Tx}}$ & 106.129 & 106.176 & 94.378 & 108.446 & 132.190 \\
			\rowcolor{gray!45}$F_{\mathrm{RcvdSuccess}}$ & 102.628 & 102.689 & 85.369 & 62.673 & 53.705\\
			\bottomrule
		\end{tabular}
		\label{table:tx_data_all_files}
		\vspace*{-\baselineskip}
	\end{subtable}
\end{table}

A comparative summary of data and file transmissions, on average, during a communication pass with fixed $y_{\textrm{min}}$ DCSM, \emph{genie I}, and \emph{static I} as well as \emph{genie II} and \emph{static II} is presented in Table~\ref{table:tx_data_all}.
On average, in order to transmit 1 bit of information, MRO transmits 2.015 bits, 2.318 bits and 2.318 bits, respectively, for the fixed \emph{static I}, DCSM and \emph{genie I} with $y_{\textrm{min}}$. Over the duration of a communication pass, on average, $D_{\textrm{Rcvd}}$ with \emph{genie I} is greater than that with DCSM, and in turn with \emph{static I} with fixed $y_{\textrm{min}}$. However, the $B_{\mathrm{Tx}}$ with \emph{static I} is about 6 Gbs more than that with \emph{genie I} and DCSM. On average, the number of files successfully received with fixed \emph{static I}, DCSM, and \emph{genie I} with $y_{\textrm{min}}$ represented by $F_{\textrm{RcvdSuccess}}^{\textrm{(sI)}}$, $F_{\textrm{RcvdSuccess}}^{\textrm{(d)}}$ and $F_{\textrm{RcvdSuccess}}^{\textrm{(gI)}}$, respectively, is 85.369, 102.628, and 102.689. It is clearly visible that with \emph{static I}, we are transmitting highest number of bits however the number of successfully received files at the receiver is lowest as compared to \emph{genie I} and DCSM. 

Compared to \emph{static I}, 17.259 and 17.32 additional files are received successfully with the DCSM and \emph{genie I} with fixed $y_{\textrm{min}}$. That is, over the same duration of a communication pass, on average, 20.217 \% and 20.288 \% of additional files are successfully received at the Earth station respectively with the DCSM and \emph{genie I} with fixed $y_{\textrm{min}}$. The $F_{\textrm{RcvdSuccess}}^{\textrm{(d)}}$ is about 99.94\% of that received with $F_{\textrm{RcvdSuccess}}^{\textrm{(gI)}}$ with fixed $y_{\textrm{min}}$. This implies that the overall channel throughput expressed in bits/s as well as successful files/s, is highest with the \emph{genie}, lowest with the \emph{static}, with DCSM closely following the \emph{genie}.

\begin{figure}[h!]
	\centering
	\includegraphics[width = 5.4in]{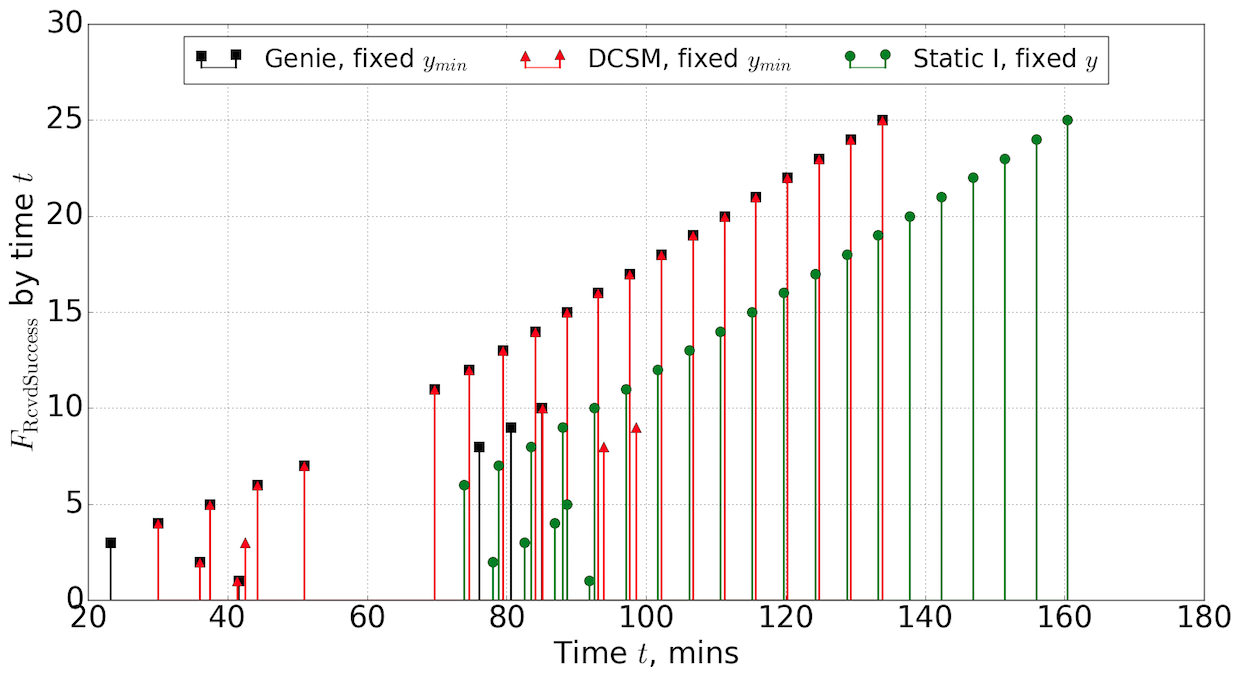}
	\caption{A plot (for first 25 files) showing the number of files successfully received at Earth station by time $t$ on a communication pass of December 9, 2001.}
	\label{fig:time_summary}
\end{figure}

Another important measure of performance is the total time required to successfully deliver a file. A plot of time required to successfully deliver first 25 files during a communication pass is presented in Fig.~\ref{fig:time_summary}. We can see that, except for some cases where delivery of file fails during its first round of transmission and needs additional round of transmissions, time required to successfully deliver a file with \emph{genie I} is comparable to that of DCSM with fixed $y_{\textrm{min}}$. However, time required for \emph{static I} with fixed $y_{\textrm{min}}$ is very large and totally different to that of both the DCSM and \emph{genie I} approach.

In general, the performance of DCSM is very close to that of {\em genie} in terms of overall channel throughput, total number of files successfully delivered to Earth station receiver, and time required to successfully deliver each file. From all the results presented here, we can clearly see that use of the {\em static} method results in a huge loss of transmission power, channel bandwidth and other associated costly resources that could be utilized efficiently if we use channel condition prediction mechanisms, vary the turbo codes dynamically and periodically use the reverse channel to send feedback on lost packets to distant spacecraft. We can see that at the cost of some minor additional bandwidth, we can achieve a very high performance gain as the amount of data loss is drastically reduced due to the use of the real time channel prediction mechanism, which can be done efficiently with the simple DCSM protocol proposed here.

\section{Conclusion}
\label{sec:conclusions}
In this paper, we presented a new content delivery protocol for deep space communications that incorporates RaptorQ codes, turbo codes, and a practical channel prediction model. We have shown that this protocol can cope with the issues of large RTT and the dynamic noise environment of space communications. We have also given an upper bound on the performance that can be achieved by incorporating real time channel condition prediction and a dynamic code rate selection strategy. At this time, use of real-time commands is very limited because of possible interaction with the planned sequences \cite{JimTaylorSeptember2006}. However, as plans are being made for the Mars landing, Mars tourism and communication with distant planets, the use of a protocol that features real-time channel condition prediction and real-time commands to dynamically control encoders used at the orbiters may provide substantial improvement in communications quality. 

\section*{Acknowledgment}
This work was supported by NASA EPSCoR program under grants NNX13AB31A and NNX14AN38A. The authors thank Kamal Oudrhiri of JPL Communications Architectures and Research Section, Radio Science Group and Philip Tsao of JPL Communications Architectures and Research Section, Signal Processing and Networks Group for helping with the AWVR data used in this manuscript. The authors also thank Philip Tsao for his helpful suggestions and comments in this work.

\bibliographystyle{ieeetr}
\bibliography{bibliographs}
\end{document}